\documentclass[12pt]{article}
\pdfoutput=1

\usepackage[a4paper,text={16.8cm,22.4cm}]{geometry}
\usepackage{amsmath,amsfonts,braket,slashed,amssymb,tikz,bm,psfrag,graphicx,color,dsfont}
\usepackage{multicol}
\usepackage{ctable}

\RequirePackage[sort&compress,square,comma,numbers]{natbib}
\allowdisplaybreaks
\begin{document}

\begin{titlepage}

\begin{flushright}
\normalsize
MITP/16-025\\ 
March 18, 2016
% arXiv:1603.05978
% v1: March 18, 2016
% v2: April 4, 2016
% v3: July 22, 2016
\end{flushright}

\vspace{1.0cm}
\begin{center}
\Large\bf\boldmath
Diphoton Resonance from a Warped Extra Dimension
\end{center}

\vspace{0.5cm}
\begin{center}
Martin Bauer$^a$, Clara H\"orner$^b$ and Matthias Neubert$^{b,c}$\\
\vspace{0.7cm} 
{\sl ${}^a$Institut f\"ur Theoretische Physik, Universit\"at Heidelberg\\
Philosophenweg 16, 69120 Heidelberg, Germany\\[3mm]
${}^b$PRISMA Cluster of Excellence \& Mainz Institute for Theoretical Physics\\
Johannes Gutenberg University, 55099 Mainz, Germany}\\[3mm]
${}^c$Department of Physics \& LEPP, Cornell University, Ithaca, NY 14853, U.S.A.\\
\end{center}

\vspace{0.8cm}
\begin{abstract}
We argue that extensions of the Standard Model (SM) with a warped extra dimension, which successfully address the hierarchy and flavor problems of elementary particle physics, can provide an elegant explanation of the 750~GeV diphoton excess recently reported by ATLAS and CMS. A gauge-singlet bulk scalar with ${\cal O}(1)$ couplings to fermions is identified as the new resonance $S$, and the vector-like Kaluza-Klein excitations of the SM quarks and leptons mediate its loop-induced couplings to photons and gluons. The electroweak gauge symmetry almost unambiguously dictates the bulk matter content and hence the hierarchies of the $S\to \gamma\gamma$, $WW$, $ZZ$, $Z\gamma$, $t\bar t$ and dijet decay rates. We find that the $S\to Z\gamma$ decay mode is strongly suppressed, such that $\mbox{Br}(S\to Z\gamma)/\mbox{Br}(S\to\gamma\gamma)<0.1$. The hierarchy problem for the new scalar boson is solved in analogy with the Higgs boson by localizing it near the infrared brane. The infinite sums over the Kaluza-Klein towers of fermion states converge and can be calculated in closed form with a remarkably simple result. Reproducing the observed $pp\to S\to\gamma\gamma$ signal requires Kaluza-Klein masses in the multi-TeV range, consistent with bounds from flavor physics and electroweak precision observables. 

Useful side products of our analysis, which can be adapted to almost any model for the diphoton resonance, are the calculation of the gluon-fusion production cross section $\sigma(pp\to S)$ at NNLO in QCD, an exact expression for the inclusive $S\to gg$ decay rate at N$^3$LO, a study of the $S\to t\bar t h$ three-body decay and a phenomenological analysis of portal couplings connecting $S$ with the Higgs field.
\end{abstract}

\end{titlepage}

\section{Introduction}

The 750~GeV excess in the diphoton invariant mass spectrum seen in the first 13~TeV data delivered by the Large Hadron Collider (LHC) \cite{ATLAS2015-81,CMS:2015dxe} could have far-reaching implications for physics beyond the Standard Model (SM). A combination of the ATLAS and CMS measurements at $\sqrt{s}=8$~TeV and 13~TeV yields \cite{Buttazzo:2015txu}
\begin{equation}\label{signal}
   \sigma(pp\to S\to\gamma\gamma) = (4.6\pm 1.2)\,\mbox{fb} \,.
\end{equation}
If this intriguing signal is confirmed to be due to new physics, minimal extensions of the SM with a single new scalar $S$ as well as many theoretically motivated ultraviolet (UV) completions are ruled out as possible explanations (see e.g.\ \cite{Staub:2016dxq}). The reason is that additional new particles with a large multiplicity or sizable couplings to $S$ have to enter the $S\to\gamma\gamma$ loop for both gluon-fusion or $b\bar b$-initiated production processes \cite{Franceschini:2015kwy}. Producing the resonance from other quark-initiated states results in a tension with 8~TeV data, while photon-induced production would require non-perturbatively large couplings \cite{Fichet:2015vvy,Csaki:2015vek}. Supersymmetric UV completions of the SM, which motivate such additional degrees of freedom, lack a neutral scalar candidate with appropriate couplings, and the full parameter space of the Minimal Supersymmetric SM is excluded as a consequence \cite{Gupta:2015zzs}. One thus has to resort to models with a low supersymmetry-breaking scale, which allow for a sgoldstino explanation \cite{Petersson:2015mkr}, or R-parity violating scenarios, in which the sneutrino can have large enough couplings to account for the excess \cite{Ding:2015rxx,Allanach:2015ixl}. Composite Higgs models predict several composite resonances that can facilitate a large diphoton branching ratio \cite{Harigaya:2015ezk,Belyaev:2015hgo,Son:2015vfl,Harigaya:2016pnu}. Neutral composite scalars, which appear in non-minimal composite Higgs models with larger coset structure, as well as the dilaton/radion have been considered as possible candidates for $S$. While theoretically motivated, the latter implies a small radius of the extra dimension in order to enhance the couplings to diphotons \cite{Cox:2015ckc}, unless the Higgs-radion mixing is tuned to a particular value \cite{Ahmed:2015uqt,Bardhan:2015hcr}. In this regard, the sgoldstino and radion explanations have similar effects on the scale of the UV completion of supersymmetric and composite Higgs theories, respectively. It is a tantalizing fact that many theoretically well-motivated, minimal extensions of the SM cannot explain the excess without such unpredicted consequences. On the other hand, several non-minimal extensions of the SM have been proposed, which can explain the diphoton excess along with other anomalies in the flavor sector and the anomalous magnetic moment of the muon \cite{Gupta:2015zzs,Aloni:2015mxa,Bauer:2015boy,Murphy:2015kag,Goertz:2015nkp,Dev:2015vjd,Hernandez:2015hrt,Belanger:2016ywb}.

In this paper we argue that Randall-Sundrum (RS) models featuring a warped extra dimension \cite{Randall:1999ee}, with all SM fields (with the possible exception of the Higgs boson) propagating in the bulk, can explain the observed excess in a natural way. We introduce a bulk scalar singlet, whose only renormalizable interactions -- with the exception of a possible Higgs portal -- are couplings to bilinears of vector-like bulk fermions. Remarkably, for ${\cal O}(1)$ couplings of this new scalar the diphoton excess is explained for Kaluza-Klein (KK) masses in the multi-TeV range without any additional model building. This mass scale is sufficiently large to avoid constraints from electroweak precision tests, flavor physics and Higgs phenomenology. Our results are largely insensitive to the parameters of the RS model, such as the five-dimensional (5D) masses of the fermions and their Yukawa couplings to the Higgs field. To good approximation the loop-induced couplings of the new resonance $S$ to diboson states just count the number of degrees of freedom propagating in the loop (times group-theory factors). We consider three implementations of the RS model with different fermion contents and present detailed predictions for the gluon-fusion production cross section $\sigma(pp\to S)$ and the rates for the decays $S\to\gamma\gamma$, $WW$, $ZZ$, $Z\gamma$, $gg$, $t\bar t$ and $t\bar t h$, all of which are found within current experimental bounds. We note in passing that our scenario is particularly well motivated if one assumes the new scalar to take on a vacuum expectation value, which generates the fermion bulk mass terms, thus providing a mechanism for the flavor-specific localization of fermions along the extra dimension \cite{Grossman:1999ra,Gherghetta:2000qt}. In this case, the bulk scalar would assume the role of the localizer field first introduced in the context of split fermion models \cite{Kaplan:2001ga}. We shall explore this intriguing possibility in future work. 

We are aware of only a few papers in which the possibility of an extra-dimensional origin of the diphoton signal has been explored. The authors of \cite{Cai:2015hzc} considered a model with a flat extra dimension. While such a framework does not address the hierarchy problem of the Higgs boson and the new scalar resonance, this work shares several technical similarities with our approach. However, the warped background of RS models makes our calculations more demanding. In \cite{Abel:2016pyc} it was assumed that the new resonance couples to the SM only via loops involving heavy vector-like leptons. In order to obtain the very large couplings required in this case \cite{Fichet:2015vvy,Csaki:2015vek}, the construction relies on more than one flat extra dimension, and only SM lepton fields are placed in the bulk. This treatment gives up the attractive possibility of understanding the flavor hierarchies from an extra-dimensional perspective. The authors found that the overlap integrals in their calculation required a cutoff, which was introduced by hand and motivated based on stringy arguments. In \cite{Arun:2015ubr,Geng:2016xin,Giddings:2016sfr} the new resonance was identified with the lowest spin-2 KK graviton in warped extra-dimension models.

This paper is organized as follows: In Section~\ref{sec:RSbasics} we briefly introduce the basic construction of warped extra-dimensional models and derive expressions for the mass and the wave-function of the bulk scalar $S$. In Section~\ref{sec:RSresults} we compute the effective Wilson coefficients parameterizing the couplings of $S$ to SM gauge bosons and top quarks by integrating out the heavy fermionic KK modes, considering both the minimal RS model as well as two different extensions with a custodial symmetry. Section~\ref{sec:pheno} deals with the phenomenology of the resonance $S$. In the context of an effective Lagrangian with local interactions of $S$ with SM fields, we first calculate the gluon-fusion production cross section $\sigma(pp\to S)$ at next-to-next-to-leading order (NNLO) in QCD perturbation theory, the inclusive $S\to gg$ decay rate at N$^3$LO, and the $S\to\gamma\gamma$, $WW$, $ZZ$, $Z\gamma$, $t\bar t$ decay rates at leading order. We then perform fits to the diphoton signal in the parameter space of the RS models, taking into account existing constraints from LHC Run~1 resonance searches. In Section~\ref{sec:portals} we study the impact of possible Higgs portal interactions of $S$ on the various branching fractions, including the $S\to hh$ signal. The three-body decay mode $S\to t\bar t h$ is studied in Section~\ref{sec:Stth}, before we conclude in Section~\ref{sec:concl}. Some technical details are relegated to two appendices.

\section{RS Model with a Bulk Scalar Field}
\label{sec:RSbasics}

We consider extensions of the SM with a warped extra dimension, described by an $S^1/Z_2$ orbifold parameterized by a coordinate $\phi\in[-\pi,\pi]$, with two branes localized on the orbifold fixed-points: the UV brane at $\phi=0$, and the infrared (IR) brane at $|\phi|=\pi$. The curvature $k$ and radius $r$ of the extra dimension are assumed to be of Planck size, $k\sim 1/r\sim M_\mathrm{Pl}$, and the metric reads \cite{Randall:1999ee}
\begin{equation}\label{metric}
   ds^2 = e^{-2\sigma(\phi)}\,\eta_{\mu\nu}\,dx^\mu dx^\nu - r^2d\phi^2 
   = \frac{\epsilon^2}{t^2} \left( \eta_{\mu\nu}\,dx^\mu dx^\nu 
    - \frac{1}{M_{\rm KK}^2}\,dt^2 \right) ,
\end{equation}
where $\sigma(\phi)=kr|\phi|$ is referred to as the warp factor. The quantity $L=\sigma(\pi)=kr\pi\sim 34$ measures the size of the extra dimension and is chosen so as to explain the hierarchy between the Planck scale and the TeV scale \cite{Goldberger:1999uk}. With the help of the curvature and the warp factor evaluated on the IR brane, $\epsilon=e^{-L}\sim 10^{-15}$, one defines the KK scale $M_{\rm KK}\equiv k\epsilon$. It sets the mass scale for the low-lying KK excitations of the model and controls the mass splitting between the KK modes. On the right-hand side of (\ref{metric}) we have introduced the dimensionless coordinate $t$ defined by $t=\epsilon\,e^{\sigma(\phi)}\in[\epsilon,1]$, which will be used throughout this work. It is related to the frequently used conformal coordinate $z$ by the rescaling $z=t/M_{\rm KK}$.

The hierarchy problem is solved by localizing the SM Higgs field on or near the IR brane, effectively cutting off UV-divergent contributions to the Higgs mass at the scale $\Lambda_{\rm IR}=\epsilon M_{\rm Pl}\sim\mbox{TeV}$ \cite{Randall:1999ee}. If gauge bosons and fermions are promoted to 5D bulk fields, the flavor problem can be addressed in a natural way by means of different localizations of the fermion zero modes along the extra dimension \cite{Grossman:1999ra,Gherghetta:2000qt}. The large hierarchies in the spectrum of fermion masses and mixing angles can then be reproduced by small variations of parameters in the underlying 5D Lagrangian \cite{Huber:2000ie}. The minimal RS model with bulk fields, which has the same gauge symmetry and matter content as the SM, is strongly constrained by electroweak precision observables \cite{Csaki:2002gy,Carena:2003fx}. A recent tree-level analysis of the $S$ and $T$ parameters yields the lower bound \cite{Malm:2013jia}
\begin{equation}\label{Mkkbound1}
   \qquad\qquad M_{\rm KK} > 4.9\,\mbox{TeV} ~~ @ ~~ 95\%~\mbox{CL} \qquad 
   \mbox{(minimal RS model)} \,.
\end{equation}
Since the masses of the low-lying KK excitations are typically several times heavier than $M_{\rm KK}$ (for example, the lightest KK gluon and photon have a mass of $2.45\,M_{\rm KK}$ \cite{Davoudiasl:1999tf}), this puts them out of the reach for discovery at the LHC. Bounds from electroweak precision observables are considerably relaxed if the electroweak sector respects the custodial symmetry present in the SM. This implies an enhanced bulk gauge group $SU(3)_c\times SU(2)_L\times SU(2)_R\times U(1)_X\times P_{LR}$, whose $SU(2)$ subgroups are broken on the IR brane via a Higgs bi-doublet according to $SU(2)_L\times SU(2)_R\to SU(2)_V$. The custodial $SU(2)_V$ symmetry in the IR protects the $T$ parameter \cite{Agashe:2003zs,Csaki:2003zu}. Boundary conditions on the UV brane break $SU(2)_R\times U(1)_X\to U(1)_Y$. The $P_{LR}$ symmetry prevents the left-handed $Zb\bar b$ coupling \cite{Agashe:2006at} and its flavor-changing counterparts \cite{Blanke:2008zb} from receiving too large corrections. As a result, the bound on the KK scale is lowered to \cite{Malm:2013jia}
\begin{equation}\label{Mkkbound2}
   \qquad\qquad M_{\rm KK} > 1.9\,\mbox{TeV} ~~ @ ~~ 95\%~\mbox{CL} \qquad 
   \text{(custodial RS model)} \,.
\end{equation}
Thorough discussions of this model containing many technical details can be found in \cite{Albrecht:2009xr,Casagrande:2010si}. In the following, we will consider two different versions of the custodial RS model: one with a symmetric implementation of the quark and lepton sectors (custodial model~I), and one in which the lepton sector is more minimal than the quark sector (custodial model~II) \cite{Hahn:2013nza}. 

Besides electroweak precision tests, RS models are constrained by flavor observables \cite{Agashe:2004cp,Csaki:2008zd,Blanke:2008zb,Albrecht:2009xr,Bauer:2009cf} and Higgs phenomenology \cite{Casagrande:2010si,Azatov:2010pf,Hahn:2013nza,Archer:2014jca,Malm:2014gha}. The most severe flavor constraint comes from $K\!-\!\bar K$ mixing \cite{Csaki:2008zd}. In the minimal model the KK scale is so high that this bound can be satisfied with a modest 25\% fine-tuning. For the lower values of the KK mass scale allowed in the custodial model, the flavor constraints can either be solved by means of a $5\!-\!10\,\%$ fine-tuning or by enlarging the strong-interaction gauge group in the bulk \cite{Bauer:2011ah}. Additional constraints arising from the phenomenology of the Higgs boson, such as its production cross section and decay rates into $\gamma\gamma$, $ZZ$ and $WW$, are more model dependent and can readily be made consistent with present data by adjusting some model parameters.

We identify the diphoton resonance with the lightest excitation of a new bulk scalar field $S(x,\phi)$, which is a singlet of the full bulk gauge group. In order to allow for a coupling of this field to the scalar density of the vector-like 5D fermion fields we need to implement $S(x,\phi)$ as an odd field on the $S^1/Z_2$ orbifold, such that $S(x,-\phi)=-S(x,\phi)$. The relevant terms in the action read 
\begin{equation}\label{eq:action}
   \int d^4x \int_{-\pi}^\pi\!d\phi\,r\,e^{-4\sigma(\phi)}\,\bigg[
    \frac{g^{MN}}{2} \left( \partial_M S \right) \left( \partial_N S \right) - \frac{\mu^2}{2}\,S^2 
    - \sum_f \Big( \mbox{sgn}(\phi)\,\bar f\,\bm{M}_f f + S\,\bar f\,\bm{G}_f f \Big) \bigg] \,,
\end{equation}
where the sum extends over all 5D fermion multiplets $f$. Even in the minimal RS model there exists a 4-component vector-like 5D fermion field for every Weyl fermion of the SM. The SM fermions correspond to the zero modes of these fields, which become massive after electroweak symmetry breaking. Consequently, for each SM fermion there exist two towers of KK excitations \cite{Casagrande:2008hr}. In extensions of the RS model with a custodial symmetry additional exotic matter fields are introduced, which have no zero modes but give rise to additional towers of KK excitations, thereby increasing the number of vector-like fermions of the model \cite{Agashe:2003zs,Csaki:2003zu}. The bulk masses $\bm{M}_f$ and couplings $\bm{G}_f$ are hermitian matrices in generation space. By means of field redefinitions one can arrange that $\bm{M}_f$ are real, diagonal matrices. From now on we will always work in this so-called bulk mass basis. The values of the bulk masses determine the profiles of the SM fermions along the extra dimension, which generically turn out to be localized near one of the two branes \cite{Grossman:1999ra,Gherghetta:2000qt}. Note that there is the intriguing possibility that the bulk masses could be generated dynamically in models where the scalar field $S$ acquires a vacuum expectation value $w$, such that $\bm{M}_f=w\bm{G}_f$. While we leave the detailed construction of such models to future work, we shall assume that the structure of the couplings $\bm{G}_f$ follows the structure of $\bm{M}_f$.

In (\ref{eq:action}) we have not considered the possibility of a portal coupling $\sim S\,|\Phi|^2$ connecting the field $S$ with the Higgs doublet. We will investigate the phenomenological impact of such a coupling on the various decay rates of the resonance $S$ in Section~\ref{sec:pheno}, finding rather strong constraints. An extra-dimensional setup, in which the Higgs sector is localized on the IR brane, where the $Z_2$-odd scalar field $S$ vanishes, might provide a dynamical explanation for the suppression of the portal interaction. We emphasize, however, that even with such sequestering a $Shh$ coupling is inevitably induced at one-loop order, since the 5D bulk fermions can mediate between the IR brane, where the Higgs field is localized, and the bulk, where the field $S$ lives. In our phenomenological analysis in Section~\ref{sec:pheno} we will therefore allow for the presence of a loop-suppressed portal interaction.

The solution of the field equations satisfied by the KK modes of the scalar field $S$ is obtained in complete analogy to the case of a bulk scalar field studied in \cite{Cacciapaglia:2006mz,Archer:2012qa,Malm:2013jia}. Imposing the KK decomposition (with $t=\epsilon\,e^{\sigma(\phi)}$)
\begin{equation}
   S(x,\phi) = \frac{e^{\sigma(\phi)}}{\sqrt{r}}\,\sum_n\,S_n(x)\,\chi_n^S(t) \,,
\end{equation}
the profile functions $\chi_n^S(t)$ are obtained from the equation of motion
\begin{equation}
   \left( t^2\partial_t^2 + t\,\partial_t - \beta^2 + t^2 x_n^2 \right) \frac{\chi_n^S(t)}{t} = 0 \,,
\end{equation}
where $x_n=m_n^S/M_{\rm KK}$ and $\beta^2=4+\mu^2/k^2$. To obtain canonically normalized kinetic terms for the KK modes, we must impose the normalization condition
\begin{equation}
   \frac{2\pi}{L} \int_\epsilon^1\!\frac{dt}{t}\,\chi_m^S(t)\,\chi_n^S(t) = \delta_{mn} \,.
\end{equation}
Requiring the Dirichlet boundary condition $\chi_n^S(\epsilon)=0$ on the UV brane, one finds the general solution
\begin{equation}
   \chi_n^S(t) = N_n\,t \left[ J_\beta(x_n t) - r_n\,J_{-\beta}(x_n t) \right] , \qquad  
   r_n = \frac{J_\beta(\epsilon x_n)}{J_{-\beta}(\epsilon x_n)} 
   \approx \frac{\Gamma(1-\beta)}{\Gamma(1+\beta)} \left( \frac{\epsilon x_n}{2} \right)^{2\beta} ,
\end{equation}
where $N_n$ is a normalization constant. In order to obtain a relatively light mass $m_S\equiv m_1^S\approx 750$~GeV for the lightest scalar resonance, we impose the mixed boundary condition $\chi_n^S(1)=\xi\,\chi_n^{S\prime}(1)$ on the IR brane, which can be engineered by adding brane-localized terms to the action. In the limits $\xi\to 0$ and $\xi\to\infty$ one recovers the special cases of the Dirichlet boundary condition $\chi_n^S(1)=0$ and the Neumann boundary condition $\chi_n^{S\prime}(1)=0$, respectively. In the general case, we obtain
\begin{equation}
    \left[ 1 - \xi(1-\beta) \right] J_\beta(x_n) - \xi\,x_n\,J_{\beta-1}(x_n)
    = r_n\,\Big\{ \left[ 1 - \xi(1+\beta) \right] J_{-\beta}(x_n) 
     - \xi\,x_n\,J_{-\beta-1}(x_n) \Big\} \,,
\end{equation}
and due to the smallness of $r_n\propto\epsilon^{2\beta}$ the right-hand side can be set to zero to excellent approximation. It follows that the mass of the lightest resonance is given by
\begin{equation}
   x_1^2\approx \frac{4(1+\beta) \left[ 1 - \xi(1+\beta) \right]}{1-\xi(3+\beta)} \,.
\end{equation}
The value $m_1^S\approx 750$~GeV can be achieved with a moderate tuning of parameters. For example, with $M_{\rm KK}=2$~TeV we need $\xi\approx 0.69$ for $\beta=0.5$, $\xi\approx 0.51$ for $\beta=1$, $\xi\approx 0.17$ for $\beta=5$ and $\xi\approx 0.09$ for $\beta=10$. The properly normalized profile function of the lightest resonance is given by (dropping irrelevant terms vanishing for $\epsilon\to 0$)
\begin{equation}
   \chi_1^S(t) = \sqrt{\frac{L(1+\beta)}{\pi}}\,t^{1+\beta}
    \left[ 1 - \frac{x_1^2}{4} \left( \frac{t^2}{1+\beta} - \frac{1}{2+\beta} \right) 
    + {\cal O}(x_1^4) \right] .
\end{equation}
The parameter $\beta$ controls the localization of the bulk scalar, and in analogy to the case of a bulk Higgs boson we will assume that $\beta>0$ (i.e., $\mu^2>-4k^2$) \cite{Breitenlohner:1982jf}. For values $\beta={\cal O}(1)$ the scalar has a wide profile along the extra dimension, while for $\beta\gg 1$ it is localized near the IR brane; in fact, we have
\begin{equation}\label{branelimit}
   \chi_1^S(t) \stackrel{\beta\to\infty}{=} 
    \sqrt{\frac{L(1+\beta)}{\pi}}\,\frac{1}{2+\beta}\,\delta(t-1) \,.
\end{equation}
While there is no particular reason why the bulk scalar should be localized near the IR brane, we will find that our results take a particularly simple form in this limit.

\section{Diboson Signals from Warped Space}
\label{sec:RSresults}

In the models we consider, the masses of the KK excitations of gauge bosons and fermions are bound by constraints from electroweak precision and flavor observables to lie in the multi-TeV range. The 750~GeV resonance is considerably lighter, and it is thus justified to integrate out the tower of fermion KK modes in computing the decays of $S$ to diboson or fermionic final states. Below the KK mass scale we define the effective Lagrangian
\begin{equation}\label{Leff}
\begin{aligned}
   {\cal L}_{\rm eff} 
   &= c_{gg}\,\frac{\alpha_s}{4\pi}\,S\,G_{\mu\nu}^a G^{\mu\nu,a}
    + c_{WW}\,\frac{\alpha}{4\pi s_w^2}\,S\,W_{\mu\nu}^a W^{\mu\nu,a}
    + c_{BB}\,\frac{\alpha}{4\pi c_w^2}\,S\,B_{\mu\nu}B^{\mu\nu} \\
   &\quad\mbox{}- \left( S\,\bar Q_L\hat{\bm{Y}}_u\,\tilde\Phi\,u_R
    + S\,\bar Q_L\hat{\bm{Y}}_d\,\Phi\,d_R + S\,\bar L_L\hat{\bm{Y}}_e\,\Phi\,e_R + \mbox{h.c.} \right) ,
\end{aligned}
\end{equation}
in which $G_{\mu\nu}^a$, $W_{\mu\nu}^a$ and $B_{\mu\nu}$ are the field strength tensors of $SU(3)_c$, $SU(2)_L$ and $U(1)_Y$, respectively, $\Phi$ is the scalar Higgs doublet, and $s_w=\sin\theta_w$ and $c_w=\cos\theta_w$ are functions of the weak mixing angle. Since the mass of the new resonance is much larger than the electroweak scale, it is appropriate to write the effective Lagrangian in the electroweak symmetric phase. Upon electroweak symmetry breaking the second and third operator in the first line generate the couplings of $S$ to pairs of electroweak gauge bosons. In particular, the resulting diphoton coupling is
\begin{equation}
   {\cal L}_{\rm eff} \ni c_{\gamma\gamma}\,\frac{\alpha}{4\pi}\,S\,F_{\mu\nu} F^{\mu\nu} \,, \quad
   \mbox{with} ~~ c_{\gamma\gamma} = c_{WW} + c_{BB} \,.
\end{equation}
The terms in the second line in (\ref{Leff}) describe the couplings of $S$ to fermion pairs (with or without a Higgs boson). In our model these couplings have a hierarchical structure, and the dominant effect by far is the coupling to the top quark. Rewriting $\mbox{Re}[(\hat Y_u)_{33}]=c_{tt}\,y_t$ (after transformation to the mass basis), where $y_t=\sqrt{2} m_t/v$ is the top-quark Yukawa coupling, we can express the corresponding term as
\begin{equation}\label{Ltth}
   {\cal L}_{\rm eff} \ni - c_{tt}\,m_t \left( 1 + \frac{h}{v} \right) S\,\bar t t + \dots \,.
\end{equation}
The Wilson coefficients in the effective Lagrangian are suppressed by the mass scale of the heavy KK particles, $c_{ii}\propto 1/M_{\rm KK}$. In the remainder of this section we will calculate these coefficients at the matching scale $\Lambda_{\rm KK}=\mbox{few}\times M_{\rm KK}$ corresponding to the masses of the low-lying KK modes, which give the dominant contributions. 

It is well known that the two-gluon operator has a non-trivial QCD evolution \cite{Inami:1982xt,Grinstein:1988wz} and mixes with the operator in (\ref{Ltth}) under renormalization \cite{Collins:1976yq}. These effects are discussed in detail in Appendix~\ref{app:RGmixing}. When the strong coupling $\alpha_s$ and the Yukawa coupling $y_t$ are factored out from the definitions of $c_{gg}$ and $c_{tt}$ as we have done above, evolution effects from the high matching scale $\Lambda_{\rm KK}$ to the scale $\mu=m_S$ only arise at NLO in renormalization-group (RG) improved perturbation theory. At this order they give rise to the simple relations
\begin{equation}
\begin{aligned}
   c_{gg}(\mu) &= \left[ 1 + \frac{\beta_1}{4\beta_0}\,
    \frac{\alpha_s(\mu)-\alpha_s(\Lambda_{\rm KK})}{\pi} \right] c_{gg}(\Lambda_{\rm KK}) \,, \\
   c_{tt}(\mu) &= c_{tt}(\Lambda_{\rm KK}) + \frac{3 C_F}{\beta_0}\,
    \frac{\alpha_s(\mu)-\alpha_s(\Lambda_{\rm KK})}{\pi}\,c_{gg}(\Lambda_{\rm KK}) \,,
\end{aligned}
\end{equation}
where $\beta_0=7$, $\beta_1=26$ and $C_F=4/3$. The coefficients $c_{WW}$ and $c_{BB}$ remain invariant under QCD evolution. Even if the coupling of $S$ to top quarks would be absent at the high matching scale, it is inevitably induced through RG evolution; however, this is a very small effect. For $\Lambda_{\rm KK}=5$~TeV and $\mu=750$~GeV we find $c_{tt}(\mu)\approx c_{tt}(\Lambda_{\rm KK})+0.0028\,c_{gg}(\Lambda_{\rm KK})$ and $c_{gg}(\mu)\approx 1.0045\,c_{gg}(\Lambda_{\rm KK})$. Higher-order QCD corrections to the Wilson coefficients at the high matching scale are likely to have a more important impact. For instance, they enhance the top-quark contribution to the Higgs-boson production cross section in the SM by about 20\% \cite{Schroder:2005hy,Chetyrkin:2005ia}. To be conservative we will not include such enhancement factors in our analysis.

\subsection{Diboson couplings induced by KK fermion exchange}

\begin{figure}
\begin{center}
\includegraphics[]{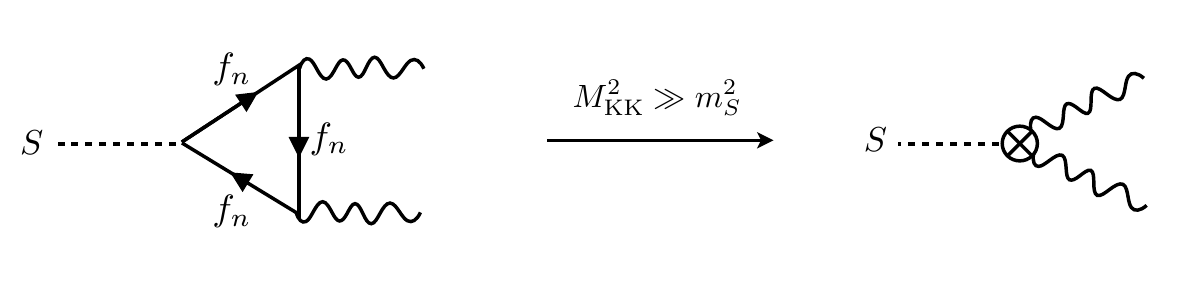}
\vspace{-2mm}
\caption{\label{fig:Effvertex} 
Loop diagrams involving the exchange of heavy KK fermions $f_n$ (left) can be described, at low energies, by effective local interactions coupling $S$ to diboson states (right).}
\vspace{-2mm}
\end{center}
\end{figure}

Since the scalar field $S$ is a gauge singlet, its couplings to gauge bosons are induced by fermion loop diagrams, such as those shown in Figure~\ref{fig:Effvertex}. The relevant couplings in (\ref{eq:action}) are parameterized by the matrices $\bm{G}_f$, while the profiles of the fermions along the extra dimension depend on the (diagonal) bulk mass matrices $\bm{M}_f$. It is conventional to define dimensionless bulk mass parameters by $\bm{c}_f=\pm\bm{M}_f/k$, where the plus (minus) sign holds for fermion fields whose left-handed (right-handed) components have even profile functions under the $Z_2$ symmetry. In the minimal RS model the $SU(2)_L$ fermion doublets have even left-handed components, while the $SU(2)_L$ fermion singlets have even right-handed components. In extensions of the RS model fields transforming as $SU(2)_L$ triplets also have even right-handed components. Using the same sign conventions, we define dimensionless couplings $\bm{g}_f$ of $S$ to fermions via
\begin{equation}\label{gfdef}
   \bm{g}_f = \pm \frac{\sqrt{k(1+\beta)}}{2+\beta}\,\bm{G}_f \,.
\end{equation}
This definition is analogous to the definition of the dimensionless Yukawa couplings in RS models with a bulk Higgs field studied in \cite{Azatov:2009na,Malm:2013jia,Archer:2014jca}. The $\beta$-dependent terms ensure that the dimensionless couplings remain well-behaved in the limit $\beta\to\infty$ of an IR brane-localized scalar field. These matrices are hermitian but, in general, not diagonal in generation space. Since with the exception of the top quark all SM fermions have masses much below the electroweak scale, the values of most of the bulk mass parameters $\bm{c}_f$ cluster near or below the critical value $-1/2$, below which the zero-mode fermion profile is localized near the UV brane. For example, a typical set of bulk mass parameters adopted in \cite{Casagrande:2008hr} ranges from $-0.74$ for $c_{u_1}$ to $-0.47$ for $c_{Q_3}$. The only exception is the parameter $c_t\equiv c_{u_3}\approx +0.34$ of the right-handed top quark, which is positive so as to realize a localization near the IR brane. In our phenomenological analysis below we will for simplicity assume that the diagonal elements of the matrices $\bm{g}_f$ all have the same sign and magnitude, with the possible exception of $g_t\equiv (g_u)_{33}$.

In close analogy with the case of the induced $hgg$ and $h\gamma\gamma$ couplings of the Higgs boson in models where all SM field propagate in the bulk, we find that the sums over the infinite towers of KK fermion states in Figure~\ref{fig:Effvertex} converge and can be calculated in closed form using 5D fermion propagators \cite{Randall:2001gb,Contino:2004vy,Carena:2004zn}. In the unbroken phase of the electroweak gauge symmetry (i.e.\ for $v=0$), there is no mixing between fermion states belonging to different multiplets of the gauge group and the fermion propagators are diagonal matrices in generation space. The Wilson coefficients are then given by sums over the contributions from the different fermion multiplets. Mixing effects induced by electroweak symmetry breaking yield corrections of order $(m_f/m_S)^2$ relative to the leading terms we will compute. Even for the top quark these corrections are at most a few percent and can safely be neglected.

The fermion representations of the custodial RS models have been discussed in detail in \cite{Agashe:2006at,Albrecht:2009xr,Casagrande:2010si,Hahn:2013nza}. We begin with a brief description of the quark sector. As a consequence of the discrete $P_{LR}$ symmetry, the left-handed bottom quark needs to be embedded in an $SU(2)_L\times SU(2)_R$ bi-doublet with isospin quantum numbers $T_L^3=-T_R^3=-1/2$. This assignment fixes the quantum numbers of the remaining quark fields uniquely. In particular, the right-handed down-type quarks have to be embedded in an $SU(2)_R$ triplet in order to obtain a $U(1)_X$-invariant Yukawa coupling. We choose the same $SU(2)_L\times SU(2)_R$ quantum numbers for all three quark generations, which is necessary to consistently incorporate quark mixing in the anarchic approach to flavor in warped extra dimensions. Altogether, there are fifteen different quark states in the up sector and nine in the down sector (for three generations). The boundary conditions give rise to three light modes in each sector, which are identified with the SM quarks. These are accompanied by KK towers consisting of groups of fifteen and nine modes of similar masses in the up and down sectors, respectively. In addition, there is a KK tower of exotic fermion states with electric charge $Q_\lambda=5/3$, which exhibits nine excitations in each KK level. In order to compute the Wilson coefficients $c_{gg}$, $c_{WW}$ and $c_{BB}$ in (\ref{Leff}) it is most convenient to decompose these multiplets into multiplets under $SU(2)_L\times U(1)_Y$. There are two $SU(2)_L$ doublets and one triplet
\begin{equation}
   \bm{c}_Q: \quad
    \left( \begin{array}{c} u_L^{(+)} \\ d_L^{(+)} \end{array} \right)_{\!\!\frac16} , \quad
    \left( \begin{array}{c} \lambda_L^{(-)} \\ u_L^{\prime\,(-)} \end{array} \right)_{\!\!\frac76} ;
   \qquad\qquad
   \bm{c}_{\tau_1}: \quad
    \left( \begin{array}{c} \Lambda_R^{\prime\,(-)} \\ U_R^{\prime\,(-)} \\ D_R^{\prime\,(-)}
    \end{array} \right)_{\!\!\frac23} ,
\end{equation}
as well as four singlets
\begin{equation}
   \bm{c}_u: \quad
    \left( u_R^{c\,(+)} \right)_{\!\frac23} ;
   \qquad\qquad
   \bm{c}_d: \quad
    \left( D_R^{(+)} \right)_{\!-\frac13} , \quad \left( U_R^{(-)} \right)_{\!\frac23} , \quad 
    \left( \Lambda_R^{(-)} \right)_{\!\frac53} .
\end{equation}
We only show the chiral components with even $Z_2$ parity; the other chiral components are odd under the $Z_2$ symmetry. The subscript denotes the hypercharge of each multiplet. The superscripts on the fields specify the type of boundary conditions they obey on the UV brane. Fields with superscript $(+)$ obey the usual mixed boundary conditions allowing for a light zero mode, meaning that we impose a Dirichlet boundary condition on the profile functions of the corresponding $Z_2$-odd fields. These zero modes correspond to the SM quarks. Fields with superscripts $(-)$ correspond to heavy, exotic fermions with no counterparts in the SM. For these states, the Dirichlet boundary condition is imposed on the $Z_2$-even fields so as to avoid the presence of a zero mode. The UV boundary conditions for the fields of opposite $Z_2$ parity are of mixed type and follow from the field equations. Above we have indicated the bulk mass parameters associated with the various multiplets.\footnote{Fields belonging to the same $SU(2)_R$ multiplet have equal bulk mass parameters. The two doublets associated with $\bm{c}_Q$ form a bi-doublet under $SU(2)_L\times SU(2)_R$, while the three singlets associated with $\bm{c}_d$ form a triplet under $SU(2)_R$.} 
The three parameters contained in the matrix $\bm{c}_{\tau_1}$ can be related to the other ones by extending the $P_{LR}$ symmetry to the part of the quark sector that mixes with the left-handed down-type zero modes \cite{Casagrande:2010si}. It then follows that $\bm{c}_{\tau_1}=\bm{c}_d$. Whether or not this equation holds turns out to be irrelevant to our discussion. 

In the custodial model~I the lepton sector is constructed in analogy with the quark sector \cite{Albrecht:2009xr}. It consists of two $SU(2)_L$ doublets and one triplet
\begin{equation}
   \bm{c}_L: \quad
    \left( \begin{array}{c} \nu_L^{(+)} \\ e_L^{(+)} \end{array} \right)_{\!\!-\frac12} , \quad
    \left( \begin{array}{c} \psi_L^{(-)} \\ \nu_L^{\prime\,(-)} \end{array} \right)_{\!\!\frac12} ;
   \qquad\qquad
   \bm{c}_{\tau_3}: \quad
    \left( \begin{array}{c} \Psi_R^{\prime\,(-)} \\ N_R^{\prime\,(-)} \\ E_R^{\prime\,(-)}
    \end{array} \right)_{\!\!0} ,
\end{equation}
as well as four singlets
\begin{equation}
   \bm{c}_\nu: \quad
    \left( \nu_R^{c\,(+)} \right)_{\!0} ;
   \qquad\qquad
   \bm{c}_e: \quad
    \left( E_R^{(+)} \right)_{\!-1} , \quad \left( N_R^{(-)} \right)_{\!0} , \quad 
    \left( \Psi_R^{(-)} \right)_{\!1} .
\end{equation}
Again we only show the chiral components with even $Z_2$ parity. There are fifteen different lepton states in the neutrino sector and nine in the charged-lepton sector. The boundary conditions give rise to three light modes in each sector, which are identified with the SM neutrinos and charged leptons. These are accompanied by KK towers consisting of groups of fifteen and nine modes in the two sectors, respectively. In addition, there is a KK tower of exotic lepton states with electric charge $Q_\psi=+1$, which exhibits nine excitations in each KK level. The three parameters contained in the matrix $\bm{c}_{\tau_3}$ can be related to the other ones by requiring an extended $P_{LR}$ symmetry, in which case $\bm{c}_{\tau_3}=\bm{c}_e$. In the custodial model~II the lepton sector is more minimal \cite{Hahn:2013nza}. It consists of one $SU(2)_L$ doublet and two singlets
\begin{equation}
   \bm{c}_L: \quad
    \left( \begin{array}{c} \nu_L^{(+)} \\ e_L^{(+)} \end{array} \right)_{\!\!-\frac12} ;
   \qquad\qquad
   \bm{c}_e: \quad
    \left( e_R^{c\,(+)} \right)_{\!-1} , \quad \left( N_R^{\prime\,(-)} \right)_{\!0} .
\end{equation}
The choice of the boundary conditions is such that the zero modes correspond to the light leptons of the SM, without a right-handed neutrino. Note that the minimal RS model is obtained by simply omitting all multiplets containing fields carrying a superscript ``$(-)$'' from the above list.

The calculation of the one-loop diagrams in Figure~\ref{fig:Effvertex} proceeds in complete analogy with the corresponding calculation for a bulk Higgs field performed in \cite{Malm:2013jia}. One evaluates the amplitude in terms of an integral over 5D propagator functions, employs the KK representation of these functions in terms of infinite sums, simplifies the resulting expression and recasts it in the form of an integral over a single 5D fermion propagator. Adapting these steps to the present case, we obtain the expressions (for $v=0$)
\begin{equation}\label{master}
\begin{aligned}
   c_{gg} &= - \sum_{f=q}\,\frac{d_f}{2} \int_0^1\!dx \int_0^1\!dy \left( 1 - 4xy\bar y \right)
    T_f(-xy\bar y m_S^2-i0) \,, \\
   c_{WW} &= - \sum_{f=q,l} N_c^f T_f \int_0^1\!dx \int_0^1\!dy \left( 1 - 4xy\bar y \right)
    T_f(-xy\bar y m_S^2-i0) \,, \\
   c_{BB} &= - \sum_{f=q,l} N_c^f d_f Y_f^2 \int_0^1\!dx \int_0^1\!dy \left( 1 - 4xy\bar y \right)
    T_f(-xy\bar y m_S^2-i0) \,,
\end{aligned}
\end{equation}
which only differ in group-theory factors. The sum in the first line runs over quark states only. Here $d_f$ is the dimension of the $SU(2)_L$ multiplet, $T_f$ is the Dynkin index of $SU(2)$ ($T_f=1/2$ for doublets, $T_f=2$ for triplets, and $T_f=0$ for singlets), $Y_f$ is the hypercharge of the multiplet, and the color factor $N_c^f$ equals 3 for quarks and 1 for leptons. The variables $x$ and $y$ (with $\bar y\equiv 1-y$) are Feynman parameters. The quantity $T_f(-p^2)$ denotes an integral over the product of mixed-chirality components of the 5D fermion propagator with momentum $p^2$ and 5D coordinates $t=t'$ with the profile of the scalar resonance $S$. Explicitly, we find in the Euclidean region\footnote{This relation holds under the assumptions that $T_f(p_E^2)$ vanishes for $p_E\to\infty$, and that $p_E\,dT_f/dp_E$ vanishes for $p_E=0$ and $p_E\to\infty$. We have checked that these conditions are satisfied in our models.} 
$p_E^2=-p^2>0$
\begin{equation}
   T_f(p_E^2) = \sqrt{\frac{\pi}{L}}\,\frac{2+\beta}{\sqrt{1+\beta}} \int_\epsilon^1\!dt\,
    \chi_1^S(t) \mbox{}\, \mbox{Tr}\bigg[ \left( \pm\bm{g}_f \right) 
    \frac{\bm{\Delta}_{LR}^f(t,t;p_E^2) + \bm{\Delta}_{RL}^f(t,t;p_E^2)}{2} \bigg] \,,
\end{equation}
where the trace is over $3\times 3$ matrices in generation space. The KK representation of the propagator functions reads
\begin{equation}\label{Deltasum}
   \bm{\Delta}_{AB}^f(t,t';p_E^2) = - \sum_n\,\frac{m_n}{p_E^2+m_n^2}\,
    {\cal F}_A^{(n)}(t)\,{\cal F}_B^{(n)\dagger}(t') \,,
\end{equation}
where the normalization of the fermion profiles ${\cal F}_A^{(n)}(t)$ with $A=L,R$ is such that \cite{Carena:2012fk}
\begin{equation}
   \int_\epsilon^1\!dt\,{\cal F}_A^{(m)\dagger}(t)\,{\cal F}_A^{(n)}(t) = \delta_{mn} \,.
\end{equation}
The zero modes are massless in the limit where $v=0$ and hence give no contribution to the result at leading order. 

Note that the sum over KK modes in (\ref{Deltasum}) is logarithmically divergent by naive power counting, since the masses of the KK modes have approximately equal spacing. Nevertheless an explicit calculation of the infinite sum leads to a finite and well-behaved answer, hinting at a non-trivial interplay of the profile functions for the various KK fermions. The calculation of the propagator functions has been discussed in detail in the literature \cite{Contino:2004vy,Carena:2004zn,Csaki:2010aj,Malm:2013jia}. It requires solving a second-order differential equation subject to appropriate boundary conditions. We obtain
\begin{equation}\label{resu1}
   \frac{\bm{\Delta}_{LR}^f(t,t;p_E^2) + \bm{\Delta}_{RL}^f(t,t;p_E^2)}{2} 
   = \pm \frac{1}{2M_{\rm KK}}\,d^{(\pm)}(\bm{c}_f,p_E,t) \,,
\end{equation}
where the overall sign is the same as that in (\ref{gfdef}), and the superscript ``$(\pm)$'' refers to the boundary conditions (normal or twisted) obeyed by the fermion multiplet $f$. The functions $d^{(\pm)}$ are diagonal matrices, whose entries depend on the bulk mass parameters. Explicitly they are given by (omitting the matrix notation for simplicity)
\begin{equation}\label{fA}
\begin{aligned}
   d^{(+)}(c_f,p_E,t) 
   &= \frac{D_2(c_f,\epsilon,t,\hat p_E)\,D_1(c_f,1,t,\hat p_E)
           + D_1(c_f,\epsilon,t,\hat p_E)\,D_2(c_f,1,t,\hat p_E)}%
          {D_2(c_f,\epsilon,t,\hat p_E)\,D_1(c_f,1,t,\hat p_E)
           - D_1(c_f,\epsilon,t,\hat p_E)\,D_2(c_f,1,t,\hat p_E)} \,, \\
   d^{(-)}(c_f,p_E,t) 
   &= \frac{D_1(-c_f,\epsilon,t,\hat p_E)\,D_1(c_f,1,t,\hat p_E)
           + D_2(-c_f,\epsilon,t,\hat p_E)\,D_2(c_f,1,t,\hat p_E)}%
          {D_1(-c_f,\epsilon,t,\hat p_E)\,D_1(c_f,1,t,\hat p_E)
           - D_2(-c_f,\epsilon,t,\hat p_E)\,D_2(c_f,1,t,\hat p_E)} \,, 
\end{aligned}
\end{equation}
where 
\begin{equation}
\begin{aligned}
   D_1(c_f,a,t,\hat p_E) &= I_{-c_f-\frac12}(a\hat p_E)\,I_{c_f-\frac12}(\hat p_E t)
    - I_{c_f+\frac12}(a\hat p_E)\,I_{-c_f+\frac12}(\hat p_E t) \,, \\
   D_2(c_f,a,t,\hat p_E) &= I_{-c_f-\frac12}(a\hat p_E)\,I_{c_f+\frac12}(\hat p_E t)
    - I_{c_f+\frac12}(a\hat p_E)\,I_{-c_f-\frac12}(\hat p_E t) \,,
\end{aligned}
\end{equation}
with $\hat p_E\equiv p_E/M_{\rm KK}$, are given in terms of modified Bessel functions. In our case these functions are evaluated (by analytic continuation to the time-like region) at momenta of order $p^2\sim m_S^2\ll M_{\rm KK}^2$, so that it is possible to expand these complicated expressions in a power series. This yields
\begin{equation}
   d^{(\pm)}(c_f,p_E,t) = k_0^{(\pm)}(c_f,t) + \hat p_E^2\,k_2^{(\pm)}(c_f,t) + {\cal O}(\hat p_E^4) \,,
\end{equation}
where
\begin{align}
   k_0^{(+)}(c_f,t) &= 1 + \frac{2F^2(c_f)}{1+2c_f} \left( t^{1+2c_f} - 1 \right) , \qquad
    k_0^{(-)}(c_f,t) = 1 \,, \nonumber\\
   k_2^{(+)}(c_f,t)  
   &= \frac{2t^2\big( 1 - t^{-1-2c_f}\big)}{1-4c_f^2}
    + 2(1-\epsilon^2)\,F^4(c_f)\,\frac{t^{1+2c_f}-1}{(1-4c_f^2)(3+2c_f)} \\
   &\quad\mbox{}- 2F^2(c_f) \left[ 
    \frac{t^2\big(2 - t^{-1-2c_f}\big)}{(1-2c_f)(1+2c_f)^2}
    - \frac{2(1+c_f)\,t^{3+2c_f}}{(1+2c_f)^2(3+2c_f)}
    - \frac{1+\epsilon^2(t^{1+2c_f}-1)}{(1-2c_f)(3+2c_f)} \right] , \nonumber\\
   k_2^{(-)}(c_f,t)  
   &= \frac{2t^2\big( 1 - t^{-1-2c_f}\big)}{1-4c_f^2}
    \left[ 1 - \left( \frac{\epsilon}{t} \right)^{1-2c_f} \right] . \nonumber
\end{align}
The quantity
\begin{equation}
   F^2(c_f) = \frac{1+2c_f}{1-\epsilon^{1+2c_f}}
\end{equation}
is the well-known zero-mode profile \cite{Grossman:1999ra,Gherghetta:2000qt}, which is exponentially small for all fermions with the exception of the right-handed top quark. We note the exact boundary values $d^{(\pm)}(c_f,p_E,1)=1$ and $d^{(\pm)}(c_f,p_E,\epsilon)=\mp 1$, from which it follows that $k_n^{(\pm)}(c_f,1)=k_n^{(\pm)}(c_f,\epsilon)=0$ for all $n\ge 2$.

\begin{figure}
\begin{center}
\includegraphics[width=0.48\textwidth]{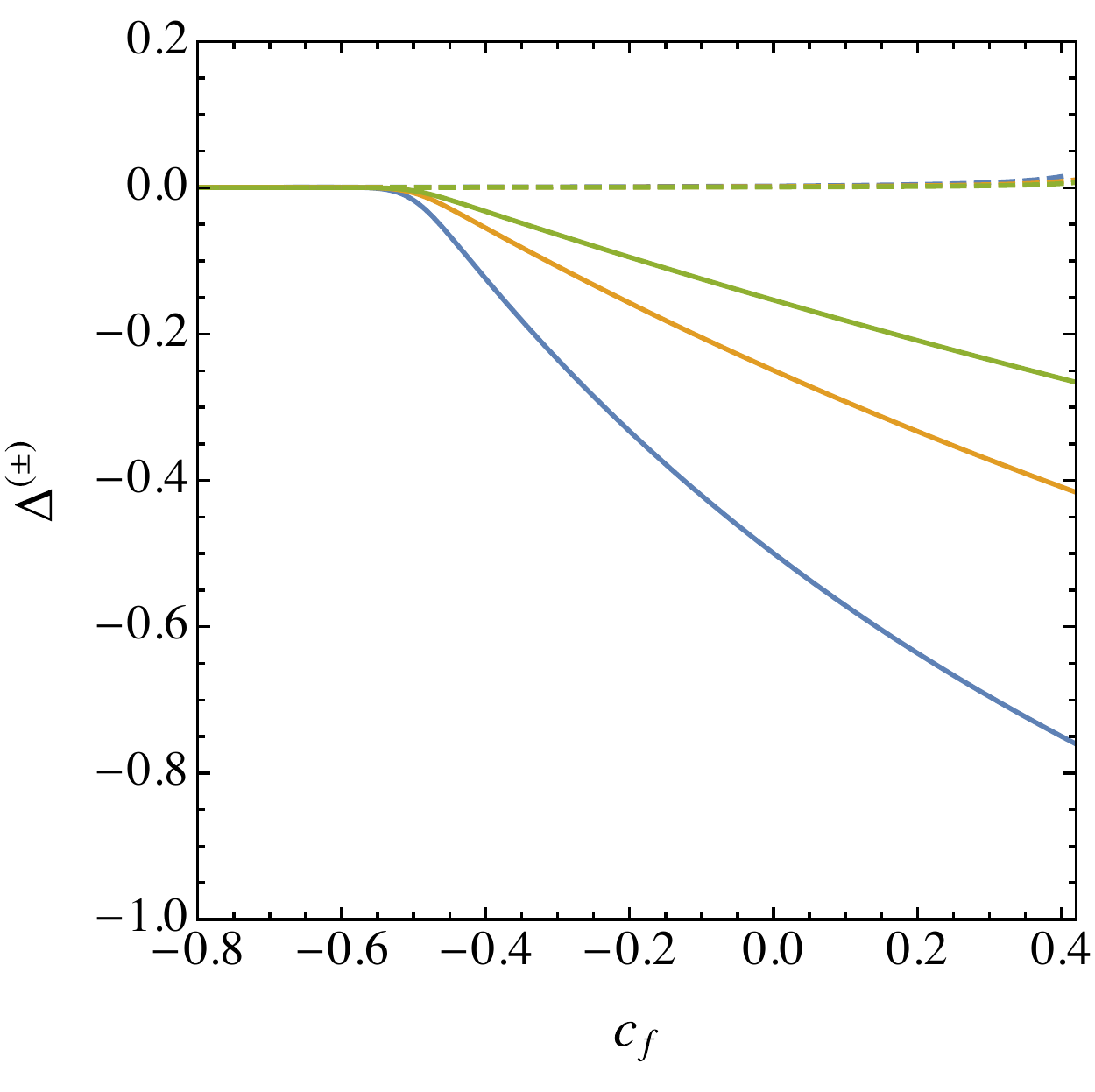}
\vspace{-2mm}
\caption{\label{fig:cpm} 
Results for the corrections $\Delta^{(+)}$ (solid) and $\Delta^{(-)}$ (dashed) for $\beta=1$ (blue), $\beta=5$ (orange) and $\beta=10$ (green) as functions of $c_f$, assuming $m_S=750$~GeV and $M_{\rm KK}=2$~TeV.}
\vspace{-2mm}
\end{center}
\end{figure}

Using these results, it follows that (taking $\epsilon\to 0$ where possible)
\begin{equation}
\begin{aligned}
   c_{gg} &= - \frac{1}{3M_{\rm KK}}\,\sum_{f=q}\,\frac{d_f}{2} 
     \int_0^1\!dt\,(2+\beta)\,t^{1+\beta} \left[ 1 - \frac{m_S^2}{4 M_{\rm KK}^2}
    \left( \frac{t^2}{1+\beta} - \frac{1}{2+\beta} \right) + \dots \right] \\
   &\qquad\mbox{}\times \mbox{Tr} \bigg[
    \bm{g}_f \left( k_0^{(\pm)}(\bm{c}_f,t)
    - \frac{7 m_S^2}{120 M_{\rm KK}^2}\,k_2^{(\pm)}(\bm{c}_f,t) + \dots \right) \bigg] \\
   &\equiv - \frac{1}{3M_{\rm KK}}\,\sum_{f=q}\,\frac{d_f}{2}\,
    \mbox{Tr} \bigg[ \bm{g}_f\,\Big( 1 + \Delta^{(\pm)}(\bm{c}_f,\beta) \Big) \bigg] \,,
\end{aligned}
\end{equation}
where (dropping irrelevant terms in $\epsilon$)
\begin{equation}
   \Delta^{(+)}(c_f,\beta) = - \frac{2F^2(c_f)}{3+\beta+2c_f}  
    + {\cal O}\bigg( \frac{m_S^2}{M_{\rm KK}^2} \bigg) \,, \qquad
   \Delta^{(-)}(c_f,\beta) = {\cal O}\bigg( \frac{m_S^2}{M_{\rm KK}^2} \bigg) \,.
\end{equation}
Analogous expressions hold for the Wilson coefficients $c_{WW}$ and $c_{BB}$, as is evident from (\ref{master}). In Figure~\ref{fig:cpm} we show the exact numerical results for $\Delta^{(\pm)}(c_f,\beta)$ as functions of the bulk mass parameter $c_f$ for various values of $\beta$. Even for $M_{\rm KK}$ as low as 2~TeV we find that the corrections of ${\cal O}(m_S^2/M_{\rm KK}^2)$ are very small and can safely be neglected. Moreover, for all fermions other than the right-handed top quark it is an excellent approximation to neglect the exponentially small quantity $F^2(c_f)$, while for the right-handed top quark we can replace $F^2(c_t)\approx 1+2c_t$. Note that in the limit $\beta\to\infty$, corresponding to a scalar resonance localized on the IR brane, we obtain the exact result $\Delta^{(\pm)}(\bm{c}_f,\beta)\to 0$, and hence the Wilson coefficients in this limit are simply given in terms of sums over the diagonal elements of the matrices $\bm{g}_f$, meaning that they essentially count the number of 5D fermionic degrees of freedom. For simplicity, we will adopt this approximation in displaying the following results. In our numerical work we will use the correct expressions, which are obtained by replacing $g_t\equiv(g_u)_{33}\to\frac{1+\beta-2c_t}{3+\beta+2c_t}\,g_t$.

We now collect our results for the Wilson coefficients in the three versions of the RS model, adopting these approximations. For the custodial model~I we find
\begin{equation}
\begin{aligned}
   c_{gg} &= - \frac{1}{3M_{\rm KK}}\,
    \mbox{Tr} \bigg( 2\bm{g}_Q + \frac12\,\bm{g}_u + \frac32\,\bm{g}_d + \frac32\,\bm{g}_{\tau_1} \bigg)
    \approx - \frac{16g_{\rm eff}}{3M_{\rm KK}} - \frac{g_t}{6M_{\rm KK}} \,, \\
   c_{WW} &= - \frac{1}{3M_{\rm KK}}\,
    \mbox{Tr} \Big( 3\bm{g}_Q + 6\bm{g}_{\tau_1} + \bm{g}_L + 2\bm{g}_{\tau_3} \Big)
    \approx - \frac{12g_{\rm eff}}{M_{\rm KK}} \,, \\
   c_{BB} &= - \frac{1}{3M_{\rm KK}}\,
    \mbox{Tr} \bigg( \frac{25}{3}\,\bm{g}_Q + \frac43\,\bm{g}_u + 10\bm{g}_d + 4\bm{g}_{\tau_1} 
    + \bm{g}_L + 2\bm{g}_e \bigg)
    \approx - \frac{236g_{\rm eff}}{9M_{\rm KK}} - \frac{4g_t}{9M_{\rm KK}} \,, 
\end{aligned}
\end{equation}
For the custodial model~II we find instead
\begin{equation}
\begin{aligned}
   c_{WW} &= - \frac{1}{3M_{\rm KK}}\,
    \mbox{Tr} \bigg( 3\bm{g}_Q + 6\bm{g}_{\tau_1} + \frac12\,\bm{g}_L \bigg)
    \approx - \frac{19g_{\rm eff}}{2M_{\rm KK}} \,, \\
   c_{BB} &= - \frac{1}{3M_{\rm KK}}\,
    \mbox{Tr} \bigg( \frac{25}{3}\,\bm{g}_Q + \frac43\,\bm{g}_u + 10\bm{g}_d + 4\bm{g}_{\tau_1} 
    + \frac12\,\bm{g}_L + \bm{g}_e \bigg)
    \approx - \frac{445g_{\rm eff}}{18M_{\rm KK}} - \frac{4g_t}{9M_{\rm KK}} \,,
\end{aligned}
\end{equation}
while $c_{gg}$ is unchanged. In the minimal RS model, the corresponding expressions read
\begin{equation}
\begin{aligned}
   c_{gg} &= - \frac{1}{3M_{\rm KK}}\,
    \mbox{Tr} \bigg( \bm{g}_Q + \frac12\,\bm{g}_u + \frac12\,\bm{g}_d \bigg)
    \approx - \frac{11g_{\rm eff}}{6M_{\rm KK}} - \frac{g_t}{6M_{\rm KK}} \,, \\
   c_{WW} &= - \frac{1}{3M_{\rm KK}}\,
    \mbox{Tr} \bigg( \frac32\,\bm{g}_Q + \frac12\,\bm{g}_L \bigg)
    \approx - \frac{2g_{\rm eff}}{M_{\rm KK}} \,, \\
   c_{BB} &= - \frac{1}{3M_{\rm KK}}\,
    \mbox{Tr} \bigg( \frac16\,\bm{g}_Q + \frac43\,\bm{g}_u + \frac13\,\bm{g}_d  
    + \frac12\,\bm{g}_L + \bm{g}_e \bigg)
    \approx - \frac{26g_{\rm eff}}{9M_{\rm KK}} - \frac{4g_t}{9M_{\rm KK}} \,. 
\end{aligned}
\end{equation}
In the last step in each line we have assumed, for simplicity, that all diagonal entries of the matrices $\bm{g}_f$ are equal to a universal value $g_{\rm eff}$. While there is no particular reason why this should be true, the near equality of all $c_f$ parameters other than $c_t$ suggests that such an approximation might be reasonable. Note that any reference to the parameter $\beta$ has disappeared, except for the small correction term multiplying $g_t$. 

In our phenomenological discussion of diboson decays in Section~\ref{sec:pheno} the ratio of the Wilson coefficients $c_{BB}$ and $c_{WW}$ will play an important role. Neglecting the small correction terms, we find $c_{BB}/c_{WW}\approx 2.19+0.04\,g_t/g_{\rm eff}$ in the custodial model~I, $c_{BB}/c_{WW}\approx 2.60+0.05\,g_t/g_{\rm eff}$ in the custodial model~II, and $c_{BB}/c_{WW}\approx 1.44+0.22\,g_t/g_{\rm eff}$ in the minimal RS model. 

\subsection{Coupling to top quarks}

The resonance $S$ has tree-level couplings to the SM fermions, which are induced after electroweak symmetry breaking. These interactions are very similar to the couplings of a bulk Higgs to fermions studied in \cite{Malm:2013jia}. The largest effects arise in the up-quark sector. We will discuss them for the case of the minimal RS model, but the final result is the same in the custodial models. Using the zero-mode profile functions derived in \cite{Casagrande:2008hr}, we find that the corresponding terms in the effective Lagrangian read (neglecting terms of order $m_h^2/M_{\rm KK}^2$)
\begin{equation}
\begin{aligned}
   {\cal L}_{\rm ferm} &= - \sum_{m,n} S(x)\,\bar u_L^{(m)}(x)\,u_R^{(n)}(x)\,
    (2+\beta) \int_0^1\!dt\,t^{1+\beta} \\
   &\quad\mbox{}\times \bigg[ x_n\,\hat a_m^{(U)\dagger}\,F(\bm{c}_Q)\,t^{\bm{c}_Q}\,
    \bm{g}_Q\,F(\bm{c}_Q)\,
    \frac{t^{1+\bm{c}_Q}-\epsilon^{1+2\bm{c}_Q}\,t^{-\bm{c}_Q}}{1+2\bm{c}_Q}\,\hat a_n^{(U)} \\
   &\hspace{1.1cm}\mbox{}+ x_m\,\hat a_m^{(u)\dagger}\,F(\bm{c}_u)\,
    \frac{t^{1+\bm{c}_u}-\epsilon^{1+2\bm{c}_u}\,t^{-\bm{c}_u}}{1+2\bm{c}_u}\,\bm{g}_u\,
    F(\bm{c}_u)\,t^{\bm{c}_u}\,\hat a_n^{(u)} \bigg] + \mbox{h.c.} \,,
\end{aligned}
\end{equation}
where $x_n=m_n/M_{\rm KK}$, and $n=1,2,3$ label the three lowest-lying states $u$, $c$ and $t$. The integrand involves a product of a $Z_2$-even fermion profile with a $Z_2$-odd one, and for the SM fermions the latter one arises from the mixing of the zero modes with their KK excitations induced by electroweak symmetry breaking. As a consequence, the overlap integrals scale with the masses of the fermions involved. The 3-dimensional vectors $\hat a_n^{(U)}$ and $\hat a_n^{(u)}$ describe the mixings in flavor space and are normalized to unity. Their entries are strongly hierarchical, with the largest entry at position $n$. The most important interaction involves the coupling of $S$ to a pair of top quarks. For the Wilson coefficient $c_{tt}$ in (\ref{Ltth}), we find to a good approximation 
\begin{equation}
\begin{aligned}
   c_{tt} &\approx \frac{1}{M_\mathrm{KK}}\,\bigg[
    \bm{g}_Q \left( 1 - \frac{F^2(\bm{c}_Q)}{3+\beta+2\bm{c}_Q} \right)
    + \bm{g}_u \left( 1 - \frac{F^2(\bm{c}_u)}{3+\beta+2\bm{c}_u} \right) \bigg]_{33} \\
   &\approx \frac{1}{M_\mathrm{KK}}\,\bigg[ \left( \bm{g}_Q \right)_{33}
    + \frac{2+\beta}{3+\beta+2c_t}\,g_t \bigg] \,.
\end{aligned}
\end{equation}

\section{Phenomenology of the Diphoton Resonance}
\label{sec:pheno}

In this section we express the production cross section and the rates for the decays of the resonance $S$ into SM particles in terms of the Wilson coefficients in the effective Lagrangians (\ref{Leff}) and (\ref{Ltth}). These results are general and can be applied to any model in which the couplings of $S$ to SM particles are induced by the exchange of some heavy new particles. We will then apply these general results to the case of the RS models studied in the previous section.

\subsection{General discussion}

At Born level, the cross section for the production of the resonance $S$ in gluon fusion at the LHC is given by
\begin{equation}\label{eq:prod}
   \sigma(pp\to S) = \frac{\alpha_s^2(\mu)\,m_S^2}{64\pi s}\,c_{gg}^2(\mu)\,
    f\hspace{-1.5mm}f_{gg}(m_S^2/s,\mu) \,,
\end{equation}
where   
\begin{equation}
   f\hspace{-1.5mm}f_{gg}(y,\mu) = \int_y^1\!\frac{dx}{x}\,f_{g/p}(x,\mu)\,f_{g/p}(y/x,\mu) 
\end{equation}
is the gluon-gluon luminosity function. The factorization and renormalization scales should be chosen of order $\mu\sim m_S$. It is well known from the analogous Higgs production cross section that higher-order QCD corrections have an enormous impact on the cross section. We have calculated these corrections up to NNLO and including resummation effects using an adaption of the public code {\tt CuTe} \cite{CuTe} developed in \cite{Becher:2011xn,Becher:2012yn}. Table~\ref{tab:sigmappS} shows our results for the ratio $\sigma(pp\to S)/c_{gg}^2(\mu)$ for the default scale choice $\mu=m_S$ and different sets of parton distribution functions (PDFs). Taking the {\tt MSTW2008} PDFs as a reference, we obtain
\begin{equation}\label{sigmas}
\begin{aligned}
   \sigma_{\rm NNLO}^{8\,{\rm TeV}}(pp\to S) 
   &= (44.9\,_{-2.4\,-2.7}^{+1.6\,+1.8})~\mbox{fb}
    \times \left( \frac{c_{gg}(m_S)}{\rm TeV} \right)^2 , \\
   \sigma_{\rm NNLO}^{13\,{\rm TeV}}(pp\to S) 
   &= (203\,_{-10\,-6}^{+~6\,+5})~\mbox{fb}
    \times \left( \frac{c_{gg}(m_S)}{\rm TeV} \right)^2 ,
\end{aligned}
\end{equation}
where the errors refer to scale variations and the variations of the PDFs. The program {\tt CuTe} also predicts the $p_T$ distribution of the produced $S$ bosons, and we find that this distribution peaks around 22~GeV. The higher-order corrections enhance the cross section by more than a factor~2 compared with the Born cross section in (\ref{eq:prod}).

\begin{table}
\centering
\begin{tabular}{c|ccccc}
\hline
$\sqrt{s}$ & {\tt MSTW2008} \cite{Martin:2009iq} & {\tt NNPDF30} \cite{Ball:2014uwa}
 & {\tt PDF4LHC15} \cite{Butterworth:2015oua} & {\tt HERA20} \cite{Cooper-Sarkar:2015boa}
  & {\tt MMHT2014} \cite{Harland-Lang:2014zoa} \\
\hline
$8~\mbox{TeV}$ & $(44.9\,_{-2.4}^{+1.6})$~fb & $(45.8\,_{-2.5}^{+1.6})$~fb
 & $(46.7\,_{-2.5}^{+1.7})$~fb & $(42.2\,_{-2.2}^{+1.4})$~fb & $(46.7\,_{-2.5}^{+1.6})$~fb \\
$13~\mbox{TeV}$ & $(203\,_{-10}^{+~6})$~fb & $(207\,_{-10}^{+~7})$~fb
 & $(208\,_{-10}^{+~7})$~fb & $(197\,_{-~9}^{+~6})$~fb & $(208\,_{-10}^{+~7})$~fb \\
\hline
\end{tabular}
\caption{\label{tab:sigmappS}
NNLO predictions for the gluon-fusion $pp\to S$ production cross section in units of $(c_{gg}/\mbox{TeV})^2$, for different sets of parton distribution functions. The quoted errors are estimated from scale variations.}
\end{table}

\begin{figure}
\begin{center}
\includegraphics[width=0.5\textwidth]{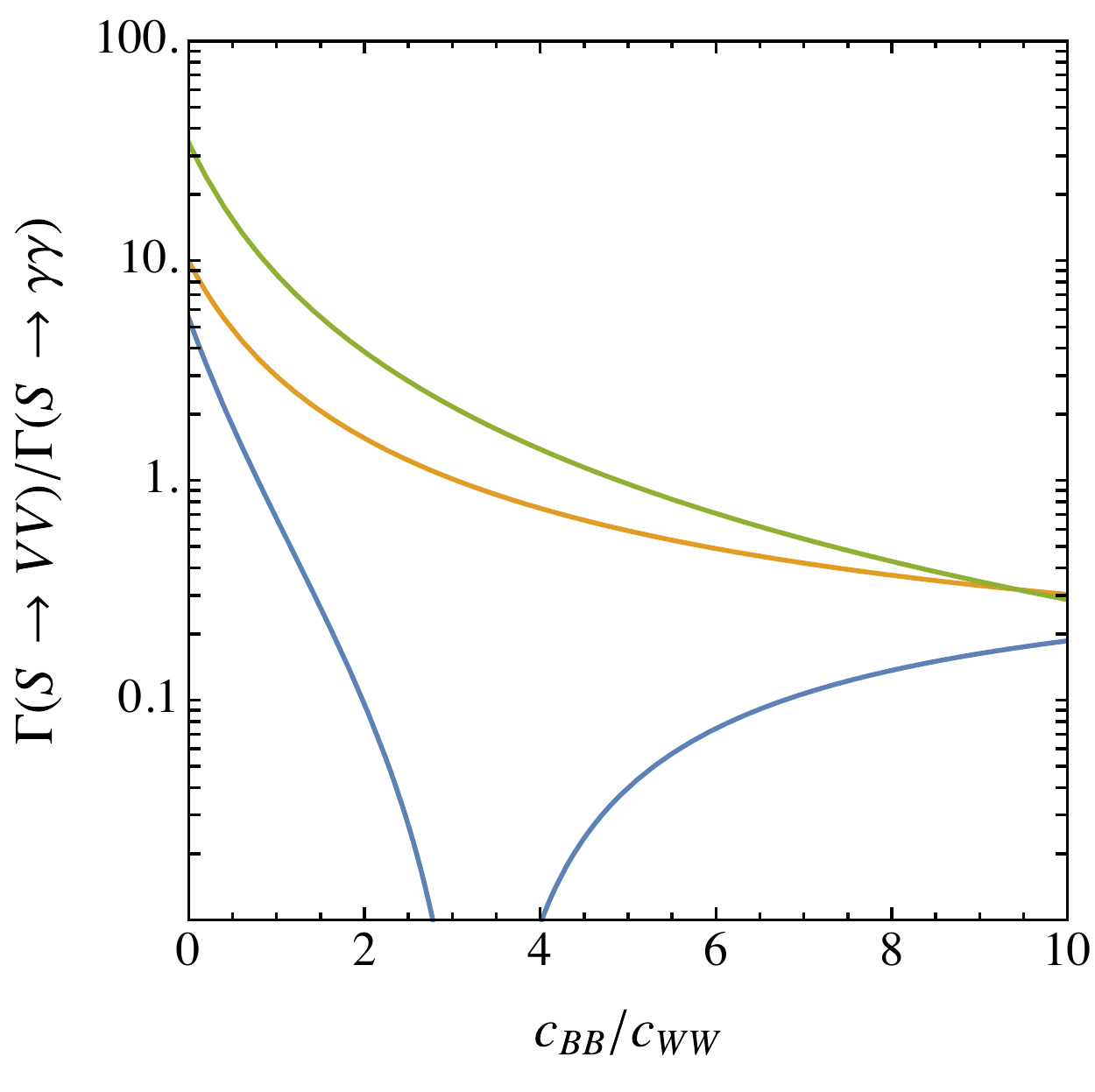}
\vspace{-2mm}
\caption{\label{fig:rates} 
Ratios of the $S\to WW$ (green), $S\to ZZ$ (orange) and $S\to Z\gamma$ (blue) decay rates to the $S\to\gamma\gamma$ rate as functions of $c_{BB}/c_{WW}$. Assuming $-2<g_t/g_{\rm eff}<1$, our calculation predicts $2.11<c_{BB}/c_{WW}<2.23$ in custodial model~I, $2.50<c_{BB}/c_{WW}<2.65$ in custodial model~II, and $1.00<c_{BB}/c_{WW}<1.66$ in the minimal RS model.}
\vspace{-2mm}
\end{center}
\end{figure}

The Wilson coefficients in (\ref{Leff}) also contribute to possible decays of the new resonance into the electroweak diboson final states $\gamma\gamma$, $WW$, $ZZ$ and $Z\gamma$, and into hadronic final states such as $gg$ and $t\bar t$. The partial decay rates for the former channels are
\begin{equation}
\begin{aligned}
   \Gamma(S\to\gamma\gamma) 
   &= \frac{\alpha^2\,m_S^3}{64\pi^3}\,\left( c_{WW} + c_{BB} \right)^2 , \\
   \Gamma(S\to WW)
   &= \frac{\alpha^2\,m_S^3}{32\pi^3}\,\frac{c_{WW}^2}{s_w^4}
    \left( 1 - 4x_W + 6x_W^2 \right) \left( 1 - 4x_W \right)^{1/2} , \\
   \Gamma(S\to ZZ)
   &= \frac{\alpha^2\,m_S^3}{64\pi^3}
    \left( \frac{c_w^2}{s_w^2}\,c_{WW} + \frac{s_w^2}{c_w^2}\,c_{BB} \right)^2
    \left( 1 - 4x_Z + 6x_Z^2 \right) \left( 1 - 4x_Z \right)^{1/2} , \\
   \Gamma(S\to Z\gamma)
   &= \frac{\alpha^2\,m_S^3}{32\pi^3}
    \left( \frac{c_w}{s_w}\,c_{WW} - \frac{s_w}{c_w}\,c_{BB} \right)^2
    \left( 1 - x_Z \right)^3 ,
\end{aligned}
\end{equation}
where $x_{W,Z}=m_{W,Z}^2/m_S^2$. While the Wilson coefficients are evaluated at the scale $\mu=m_S$, the gauge couplings and Weinberg angle are evaluated at the scale appropriate for the final-state bosons. We use $\alpha(m_Z)=1/127.94$ for $Z$ and $W$ bosons, $\alpha=1/137.04$ for the photon, and $s_W^2=0.2313$ for the weak mixing angle. Apart from known quantities, the two Wilson coefficients $c_{WW}$ and $c_{BB}$ determine these four rates entirely. It follows that any ratio of two rates is a function of the ratio $c_{BB}/c_{WW}$, which in turn is characteristic for the model under investigation. This is illustrated in Figure~\ref{fig:rates}. Note that the $S\to Z\gamma$ decay rate in particular strongly depends on the value of $c_{BB}/c_{WW}$. In the RS models considered in the previous section this decay mode turns out to be strongly suppressed.

The partial rates for decays into hadronic final states are 
\begin{equation}
\begin{aligned}
   \Gamma(S\to gg) 
   &= \frac{\alpha_s^2(\mu)\,m_S^3}{8\pi^3}\,K_{gg}(\mu)\,c_{gg}^2(\mu) \,, \\
   \Gamma(S\to t\bar t)
   &= \frac{3m_t^2(\mu)\,m_S}{8\pi}\,c_{tt}^2(\mu) \left( 1 - 4x_t \right)^{3/2} ,
\end{aligned}
\end{equation}
where all running quantities should be evaluated at $\mu\approx m_S$. In the second case $m_t(\mu)$ is the running top-quark mass (we use $m_t(m_S)=146.8$~GeV), whereas the mass ratio $x_t=m_t^2/m_S^2$ entering the phase-space factor involves the top-quark pole mass $m_t=173.34$~GeV. In many scenarios the dijet decay mode $S\to gg$ is the dominant decay channel and hence enters in the calculation of the branching fractions for all other decay modes. It is therefore important to calculate this partial rate as accurately as possible. Using existing calculations of the Higgs-boson decay rate $\Gamma(h\to gg)$ up to ${\cal O}(\alpha_s^5)$ in the heavy top-quark limit \cite{Baikov:2006ch,Schreck:2007um} it is possible to derive an exact expression for the $S\to gg$ decay rate to the same accuracy. This is discussed in detail in Appendix~\ref{app:Sgg}. We find that the impact of radiative corrections is significantly smaller than in the Higgs case, and that the perturbative series at $\mu=m_S$ exhibits very good convergence. We obtain $K_{gg}^{\rm N^3LO}(m_S)\approx 1.348$. 

\subsection{\boldmath $S$-boson phenomenology in RS models}

\begin{table}
\centering
\begin{tabular}{cccccc}
\hline
$jj$ & $WW$ & $ZZ$ & $Z\gamma$ & $t\bar t$ & $hh$ \\
\hline
$<2.5$~pb \cite{CMS:2015neg} & $<40$~fb \cite{Aad:2015agg} & $<12$~fb \cite{Aad:2015kna}
 & $<4$~fb \cite{Aad:2014fha} & $<700$~fb \cite{Aad:2015fna} & $<50$~fb \cite{CMS:2014eda} \\
\hline
\end{tabular}
\caption{\label{tab:bounds}
Bounds (at 95\% CL) on the $pp\to S\to XX$ production cross sections obtained in dijet, diboson and $t\bar t$ resonance searches performed in Run~1 of the LHC ($\sqrt{s}=8$~TeV).}
\end{table}

\begin{figure}
\begin{center}
\hspace{-.75cm}
\begin{tabular}{cc}
\includegraphics[scale=.65]{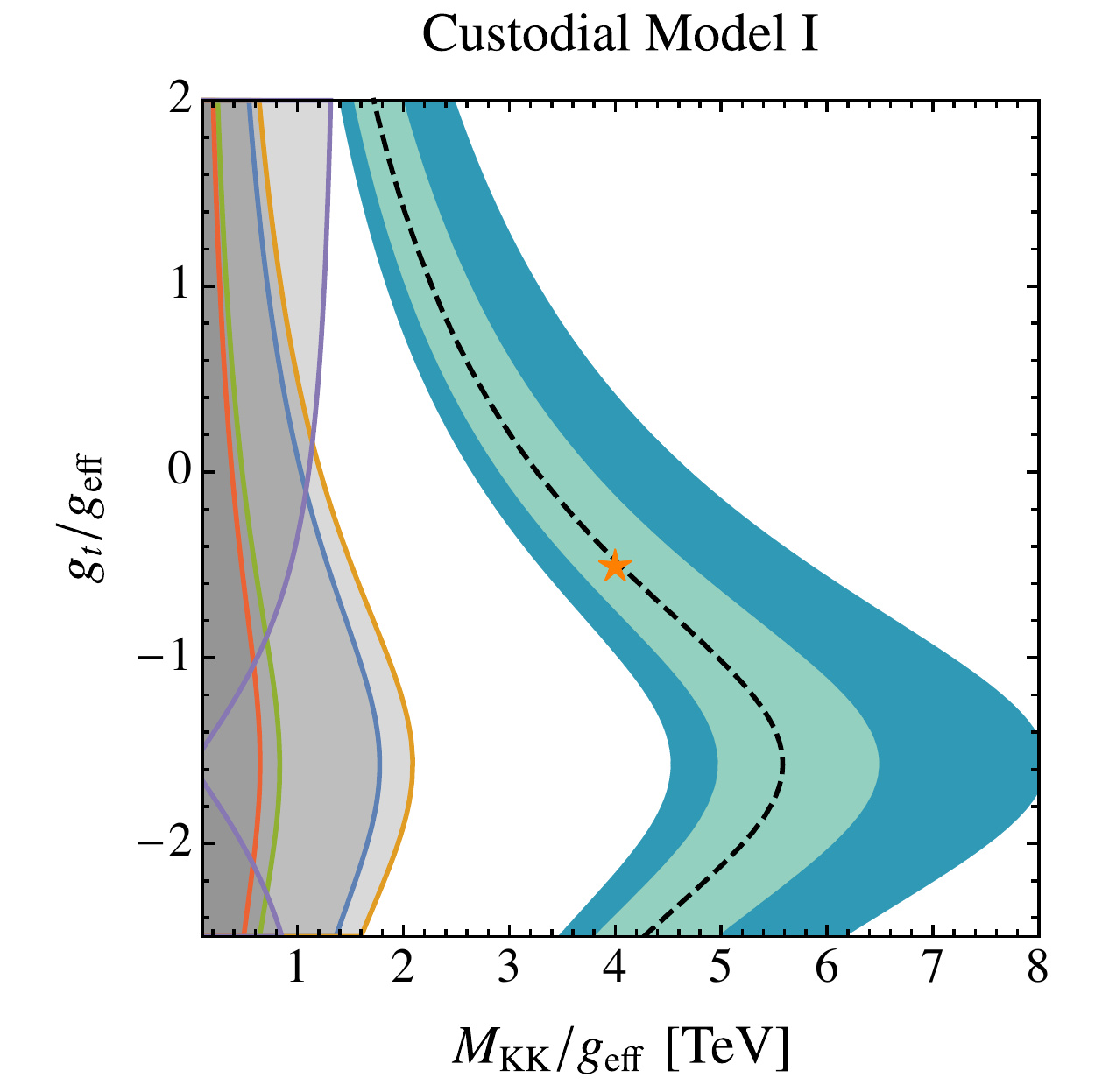} & \includegraphics[scale=.65]{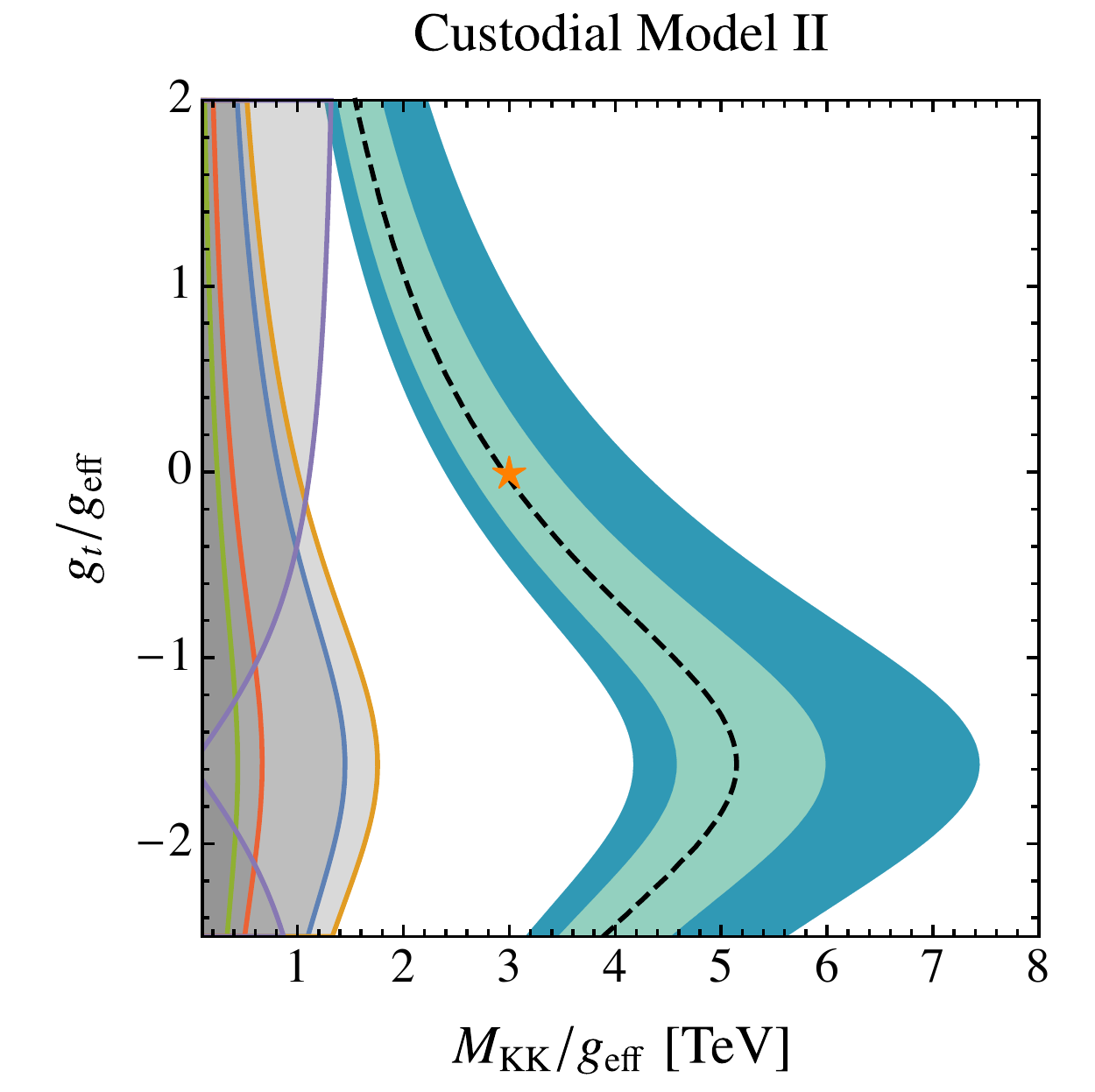} \\[4mm]
\includegraphics[scale=.65]{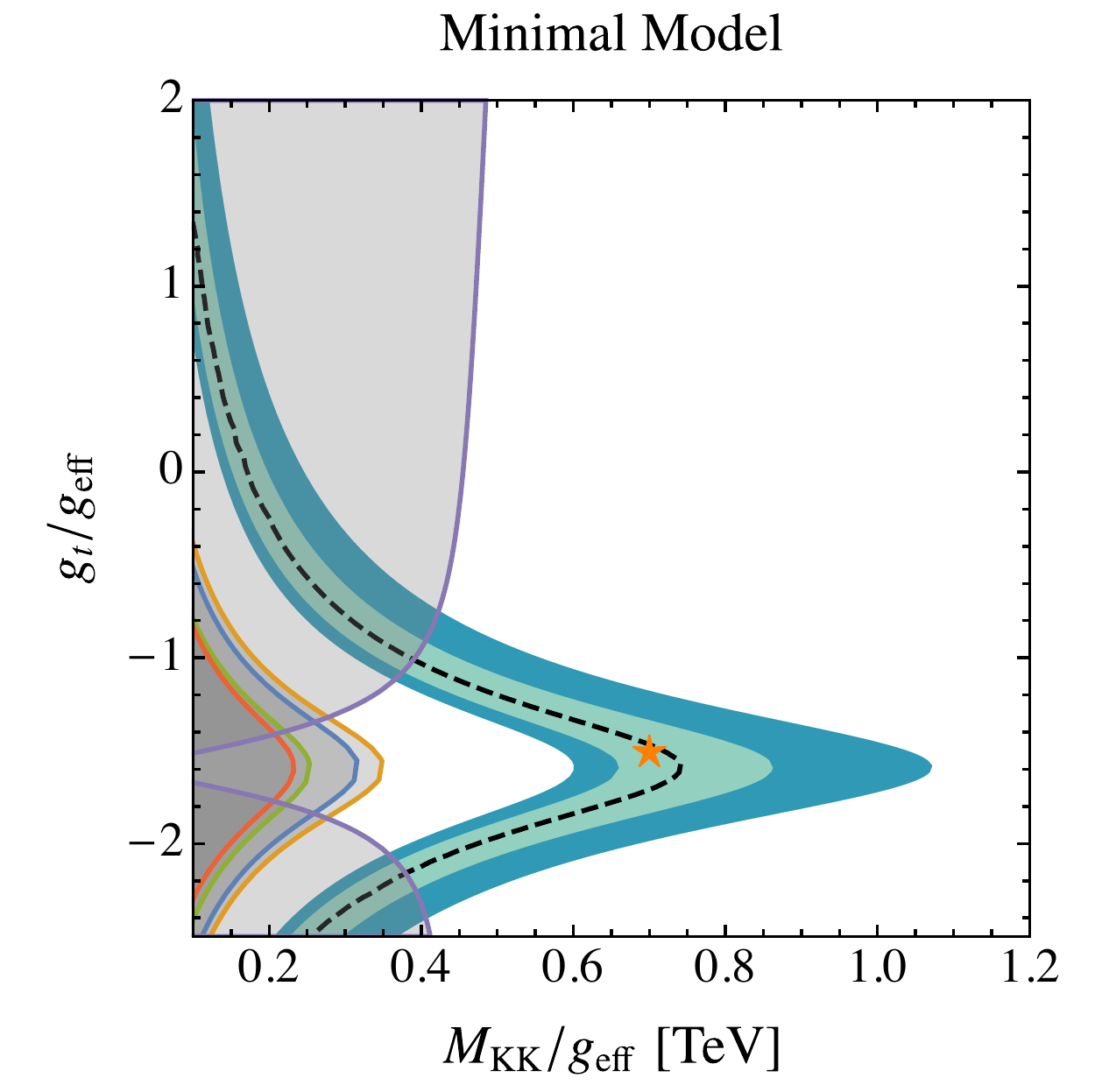} & \includegraphics[scale=.65]{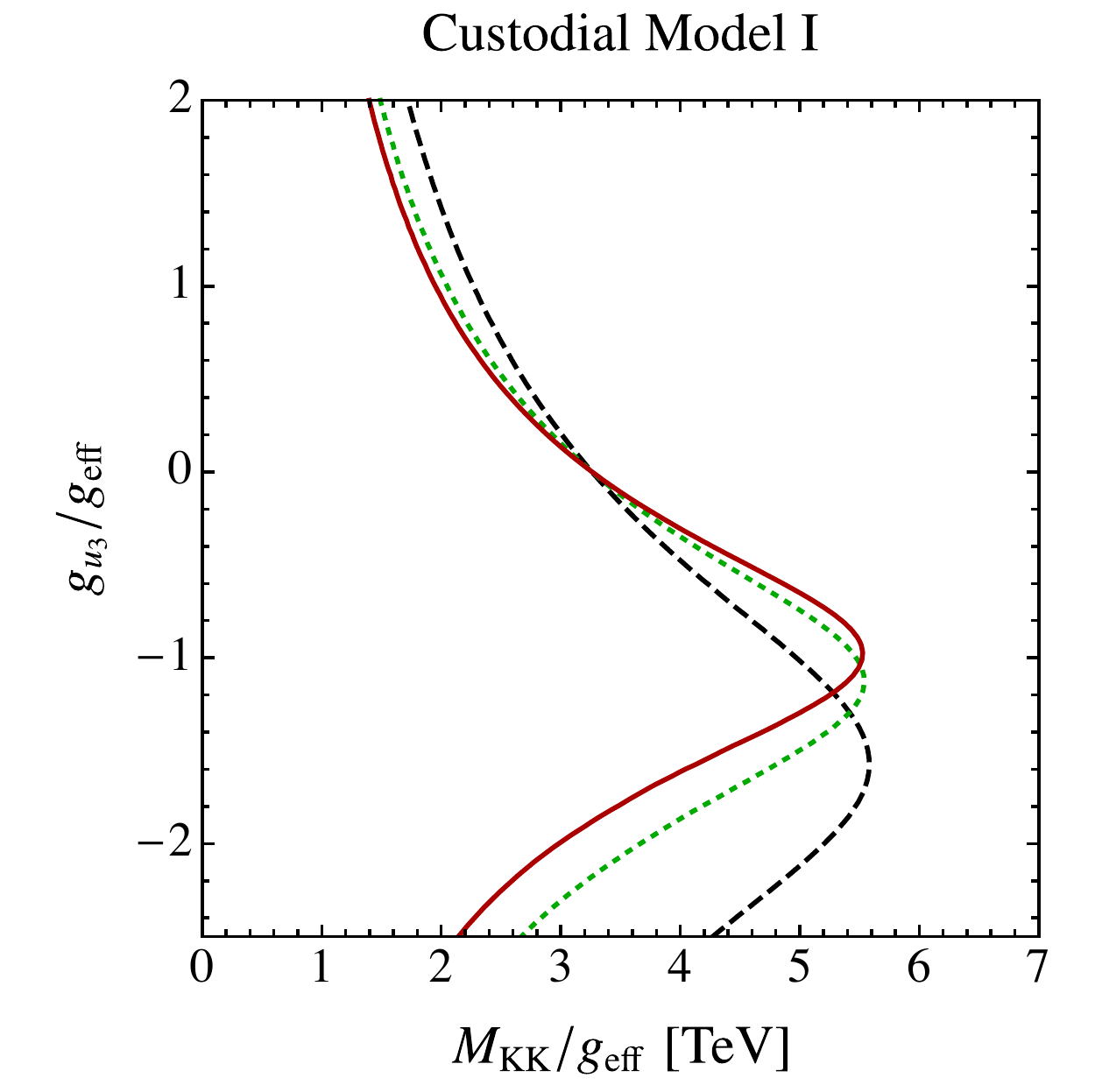}
\end{tabular}
\caption{\label{fig:plots} 
Regions in the parameter space of the RS models in which the diphoton signal is reproduced at $1\sigma$ (light blue) and $2\sigma$ (dark blue). The black dashed line corresponds to the central value shown in (\ref{signal}). The two upper panels refer to the custodial models~I and II, while the lower left panel refers to the minimal RS model. Regions excluded by bounds from resonance searches in Run~1 data (at 95\% CL) are shaded gray with boundaries drawn in red (dijets), purple ($t\bar t$), blue ($WW$), orange ($ZZ$) and green ($Z\gamma$). We use $\beta=1$ and $c_t=0.4$. The lower right panel shows the variation of the central fit result with the localization parameter $\beta$ of the scalar profile, for $\beta=1$ (black dashed), $\beta=10$ (dotted green) and $\beta=\infty$ (red).}
\end{center}
\end{figure}

We are now ready to explore the phenomenological consequences of our calculations. The challenge is to reproduce the observed diphoton rate in (\ref{signal}), while at the same time respecting existing bounds on dijet, diboson and $t\bar t$ resonance searches\footnote{Note that the reported bound for the $S\to t\bar t$ channel might in fact be considerably weaker due to interference effects not considered in the experimental analyses \cite{UliHaisch}. The potential impact of these effects has first been pointed out in \cite{Dicus:1994bm} and has recently been reemphasized in \cite{Craig:2015jba, Gori:2016zto}.}
from Run~1 of the LHC, which are collected in Table~\ref{tab:bounds}. Assuming that the new resonance $S$ is predominantly produced via gluon fusion, as is the case in the RS models we study, the corresponding bounds at $\sqrt{s}=13$~TeV are obtained by multiplying these numbers with the boost factor~4.52 corresponding to the ratio of the production cross sections in (\ref{sigmas}).

In Figure~\ref{fig:plots}, we present plots in the $M_{\rm KK}/g_{\rm eff}-g_t/g_{\rm eff}$ plane showing the $1\sigma$ and $2\sigma$ fit regions to the diphoton excess in light and dark blue. The central fit values are shown by the dashed black line. Regions excluded by bounds from resonance searches in data collected during the 8~TeV run of the LHC are shaded gray with a boundary drawn in red (dijet searches), purple ($t\bar t$ searches), blue ($WW$ searches), orange ($ZZ$ searches) and green ($Z\gamma$ searches). Throughout we use $\beta=1$ and $c_t=0.4$ for the parameters entering the contribution from the $SU(2)_L$-singlet top quark. The lower right plot shows the variation of the central fit result for different values of the localization parameter $\beta$, namely $\beta=1$ (black dashed), $\beta=10$ (dotted green) and $\beta=\infty$ (red). It is apparent that there is only a minor dependence on this parameter. Changing $c_t$ does not substantially alter the fit either. 

We observe that in the two versions of the RS model with a custodial symmetry the diphoton signal can be reproduced over a wide range of parameters without any fine tuning and without violating any of the bounds from other searches. Depending on the choice of $g_t/g_{\rm eff}$ one obtains values for $M_{\rm KK}/g_{\rm eff}$ in the range between 2 and 8~TeV. If the KK scale $M_{\rm KK}$ is close to the lower bound (\ref{Mkkbound2}) allowed by electroweak precision tests, this requires couplings $g_{\rm eff}$ in the range 0.25 to 1, which are well inside the perturbative region. In this scenario some of the low-lying KK excitations could have masses around 4~TeV, in which case they might be discovered in Run~2 of the LHC. If the KK mass scale is significantly higher a direct discovery of KK excitations will not be possible at the LHC. Nevertheless, even for $M_{\rm KK}\approx 5$~TeV (implying KK resonance masses near 10~TeV) the diphoton signal can be explained with a modest coupling $g_{\rm eff}\sim 1$. In the minimal RS model the parameter space in which the diphoton signal can be explained is more constrained. We find values in the range $M_{\rm KK}/g_{\rm eff}\sim 0.4\!-\!1$, which for a KK scale as high as the bound (\ref{Mkkbound1}) enforced by electroweak precision tests requires large couplings $g_{\rm eff}\sim 5\!-\!12$, close to the perturbativity limit. One also needs to require that the ratio $g_t/g_{\rm eff}$ is negative so as to avoid the strong constraint from $t\bar t$ resonance searches (see, however, footnote~3).

\begin{table}
\begin{center}
\begin{tabular}{l|cccccccc|c}
\hline
Br$(S\to XX)$ & $gg$ & $\gamma\gamma$ & $WW$ & $ZZ$ & $Z\gamma$ & $t\bar t$ & $hh$ & $t\bar t h$ 
 & $\Gamma_{\rm tot}$ \\
\hline
Custodial~I & 43.0\% & 1.30\% & 5.1\% & 2.1\% & 0.10\% & 47.9\% & 0 & 0.50\% & 0.08~GeV \\
Custodial~II & 28.4\% & 0.68\% & 2.1\% & 0.9\% & 0.02\% & 67.2\% & 0 & 0.70\% & 0.22~GeV \\
Minimal & 89.2\% & 0.37\% & 2.7\% & 1.0\% & 0.16\% & ~6.6\% & 0 & 0.07\% & 0.14~GeV \\
\hline
Custodial~I & 32.2\% & 0.97\% & 9.9\% & 4.6\% & 0.08\% & 48.5\% & 3.1\% & 0.60\% & 0.11~GeV \\
Custodial~II & 24.1\% & 0.58\% & 4.3\% & 2.0\% & 0.01\% & 66.9\% & 1.3\% & 0.77\% & 0.25~GeV \\
Minimal & 78.0\% & 0.32\% & 6.3\% & 2.8\% & 0.14\% & 10.2\% & 2.1\% & 0.14\% & 0.16~GeV \\
\hline
Custodial~I & 21.5\% & 0.65\% & 18.0\% & 8.7\% & 0.05\% & 42.1\% & 8.4\% & 0.59\% & 0.16~GeV \\
Custodial~II & 19.2\% & 0.46\% & ~9.1\% & 4.4\% & 0.01\% & 61.9\% & 4.2\% & 0.77\% & 0.32~GeV \\
Minimal & 60.4\% & 0.25\% & 13.7\% & 6.5\% & 0.11\% & 12.3\% & 6.5\% & 0.21\% & 0.21~GeV \\
\hline
\end{tabular}
\end{center}
\caption{\label{tab:BRs}
Branching ratios for various decay modes of the resonance $S$ in the three RS models and for the benchmark parameter points defined in (\ref{benchmark}). In the center and lower portions of the table we show the branching ratios in the presence of a small portal coupling $\lambda_1=0.02$ and 0.04, respectively, see Section~\ref{sec:portals}. The small contributions to the $S\to hh$ and $S\to t\bar th$ branching ratios resulting from the portal coupling $\lambda_2$ in (\ref{Shhrate}) and (\ref{eq52}) have been set to~0.}
\end{table}

We find it useful to define a benchmark point for each model and study the individual branching fractions for the various $S$ decay modes for these points. Specifically, we choose the points indicated by the orange stars in Figure~\ref{fig:plots}, for which (with $\beta=1$ and $c_t=0.4$)
\begin{equation}\label{benchmark}
\begin{aligned}
   & \text{Minimal model}: & M_\text{KK}/g_\text{eff} = 0.7~\mbox{TeV}, \qquad
    & g_t/g_\text{eff} = -1.5 \,, \\
   & \text{Custodial model~I}: \quad & M_\text{KK}/g_\text{eff} = 4.0~\mbox{TeV}, \qquad
    & g_t/g_\text{eff} = -0.5 \,, \\
   & \text{Custodial model~II}: & M_\text{KK}/g_\text{eff} = 3.0~\mbox{TeV}, \qquad
    & g_t/g_\text{eff} = 0 \,.
\end{aligned}
\end{equation}
In the upper portion of Table~\ref{tab:BRs} we collect the branching ratios into the various final states for these benchmark models. Note that the $S\to t\bar t$ decay rate is only calculated at lowest order in QCD and hence afflicted with some uncertainty. The $S\to t\bar t$ branching ratio is rather sensitive to the choice of $g_t/g_{\rm eff}$, while the remaining branching fractions only mildly depend on this parameter. The three-body decay mode $S\to t\bar t h$ will be discussed in Section~\ref{sec:Stth}. In the last column we show the total decay width of $S$, which is very small in our models. Given the existing Run~1 dijet bound shown in Table~\ref{tab:bounds}, it is impossible to obtain a total width exceeding a few GeV in any model in which the decay $S\to gg$ has a significant branching ratio. This is below the experimental resolution of approximately 10~GeV on $m_{\gamma\gamma}$. In our framework we can therefore not accommodate the best fit value $\Gamma_{\rm tot}\approx 45$~GeV reported by ATLAS \cite{ATLAS2015-81}. Rather, the numbers shown in the table correspond to values $\Gamma_{\rm tot}/m_S\approx (1.1\!-\!4.3)\cdot 10^{-4}$. We recall, however, that the large width $\Gamma_{\rm tot}\approx 0.06\,m_S$ is only slightly preferred by the ATLAS analysis, leading to an improvement of the fit by $0.3\sigma$ over a narrow-width scenario. An independent analysis in \cite{Buckley:2016mbr} concludes that the large-width scenario is disfavored by a combination of the ATLAS and CMS analyses of the 13~TeV data, and only slightly preferred taking into account the 8~TeV data, because it is easier to absorb the signal of a broad resonance in the background model (the local significance changes at most by $0.5\sigma$ between these options).\footnote{After the submission of our paper, CMS has reported a combined analysis of the 8~TeV and 13~TeV data using three templates with $\Gamma_{\rm tot}/m_S=1.4\cdot 10^{-4}$, $1.2\cdot 10^{-2}$ and $5.6\cdot 10^{-2}$ [CMS Collaboration, CMS-PAS-EXO-16-018]. The best fit is obtained for the narrow-width assumption with $\Gamma_{\rm tot}/m_S=1.4\cdot 10^{-4}$. A value of this order is indeed predicted in our models.}

We observe that there are rather striking differences between the three RS models considered here, even though any of the three benchmark points reproduces the diphoton signal and is consistent with all other bounds. In particular, the $S\to WW$, $ZZ$ and $t\bar t$ branching ratios vary significantly from one model to another, indicating that future measurements of these modes will provide very interesting clues about the underlying model. Note also that in all cases we find that the $S\to Z\gamma$ branching fraction is very small, so that it will be challenging to observe this mode in our scenarios. On the other hand, {\em not\/} seeing the $S\to Z\gamma$ signal would be as important a finding as seeing it.

\section{Impact of Higgs portal couplings}
\label{sec:portals}

\begin{figure}
\begin{center}
\includegraphics[width=0.75\textwidth]{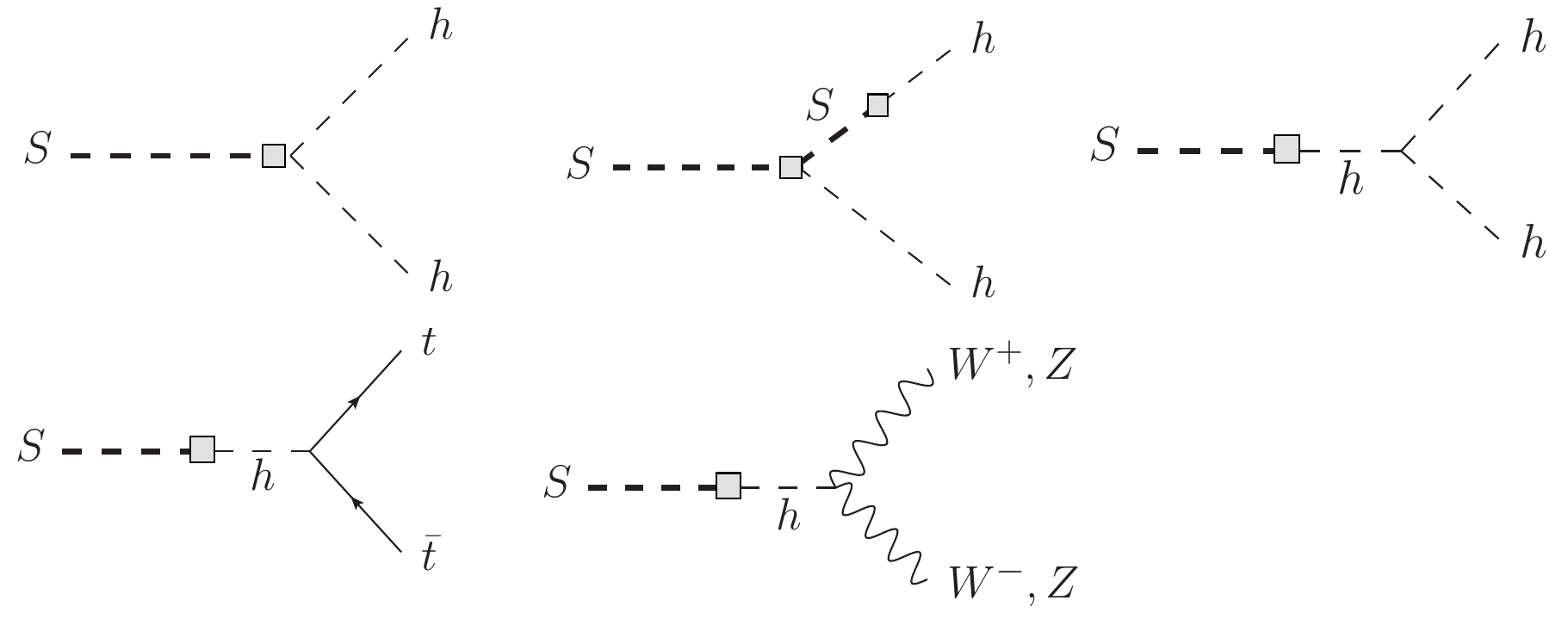}
\vspace{-2mm}
\caption{\label{fig:Shmixing} 
Diagrams contributing to $S$ decays induced by mixing with the Higgs boson.}
\vspace{-2mm}
\end{center}
\end{figure}

The most general renormalizable Lagrangian includes besides the operators in (\ref{Leff}) potential Higgs portal interactions (see e.g.\ \cite{Franceschini:2015kwy,Gupta:2015zzs,Cheung:2015cug,D'Eramo:2016mgv})
\begin{equation}\label{Leffportals}
   \delta{\cal L}_{\rm eff} 
   = - \lambda_1\,m_S\,S \left|\Phi\right|^2 - \frac{\lambda_2}{2}\,S^2 \left|\Phi\right|^2
   \ni - \frac{\lambda_1}{2}\,m_S\,S \left(v+h\right)^2 - \frac{\lambda_2}{4}\,S^2 \left(v+h\right)^2\,.
\end{equation}
In RS models the couplings $\lambda_1$ and $\lambda_2$ can be suppressed at tree level by localizing the Higgs sector on or near the IR brane, where the $Z_2$-odd bulk field for the resonance $S$ vanishes. However, starting at one-loop order the portal couplings will be induced through diagrams analogous to that shown in Figure~\ref{fig:Effvertex}, but with the external gauge fields replaced by Higgs bosons. Note that this diagram exists even if only the ``right-chirality'' couplings of the Higgs bosons are included. Also, below the electroweak scale the effective Lagrangian (\ref{Leff}) gives rise to a contribution to the portal coupling $\lambda_1$ proportional to $c_{tt}$ from top-quark loop graphs. RS models thus provide a rationale for why the portal interactions should be suppressed (by small overlap integrals or a loop factor), but it would be unjustified to omit them altogether.

After electroweak symmetry breaking, the first portal interaction gives rise to three interesting (and potentially dangerous) effects. First, the presence of a tadpole for the field $S$ requires that we define the physical field by the shift $S\to S-(\lambda_1 v^2)/(2m_S)$. Performing this shift in the Lagrangian (\ref{Leff}) generates corrections to the SM Yukawa couplings and wave-function corrections to the gauge fields. Otherwise these corrections do not have observable effects. Second, there is a tree-level decay $S\to hh$, generated by the upper two diagrams shown in Figure~\ref{fig:Shmixing}, whose decay rate (at leading order in the portal coupling $\lambda_1$) is given by 
\begin{equation}\label{Shhrate}
   \Gamma(S\to hh) = \frac{m_S}{32\pi}\,\lambda_1^2
    \left( 1 + \frac{3m_h^2-2\lambda_2 v^2}{m_S^2-m_h^2} \right)^2 \sqrt{1-4x_h} \,.
\end{equation}
Third, there is a mass mixing between the scalar resonance $S$ and the Higgs boson. At leading order in the portal coupling $\lambda_1$ this gives rise to tree-level contributions to the $S\to t\bar t$, $WW$, $ZZ$ decay amplitudes induced by the lower two diagrams in Figure~\ref{fig:Shmixing}. The corrected expression for the $S\to t\bar t$ decay width is then given by
\begin{equation}
   \Gamma(S\to t\bar t)
   = \frac{3m_t^2}{8\pi m_S} \left( 1 - 4x_t \right)^{3/2} 
    \left( m_S\,c_{tt} + \frac{\lambda_1}{1-x_h} \right)^2 .
\end{equation}
The corrected expression for the $S\to WW$ decay rate reads
\begin{align}
   \Gamma(S\to WW)
   &= \frac{m_S}{16\pi}\,\sqrt{1-4x_W}\,\bigg[
    \frac{m_S^2\,c_{WW}^2}{2} \left( \frac{\alpha}{\pi\,s_W^2} \right)^2 (1-4x_W+6x_W^2) \\
   &\quad\mbox{}+ \left( \frac{\lambda_1}{1-x_h} \right)^2 (1-4x_W+12x_W^2) 
    - 6 m_S\,c_{WW}\,\frac{\alpha}{\pi\,s_W^2}\,\frac{\lambda_1}{1-x_h}\,
    x_W(1-2x_W) \bigg] \,, \nonumber
\end{align}
and similarly for $S\to ZZ$. The mixing between the fields $S$ and $h$ also has an impact on the properties of the Higgs boson. The physical Higgs boson is given by the combination $(\cos\theta\,h-\sin\theta\,S)$, where $\sin2\theta=2\lambda_1 m_S v/(m_S^2-m_h^2)\approx 0.67\lambda_1$, with $m_S$ and $m_h$ refering to the physical masses after the field redefinitions. Existing measurements of the Higgs branching fractions constrain $\cos\theta$ to be larger than 0.86 at 95\% CL \cite{Carmi:2012in,Cheung:2015dta}, which implies the rather weak constraint $|\lambda_1|<1.3$ in our model.

\begin{figure}[t]
\begin{center}
\hspace{-.75cm}
\includegraphics[scale=.65]{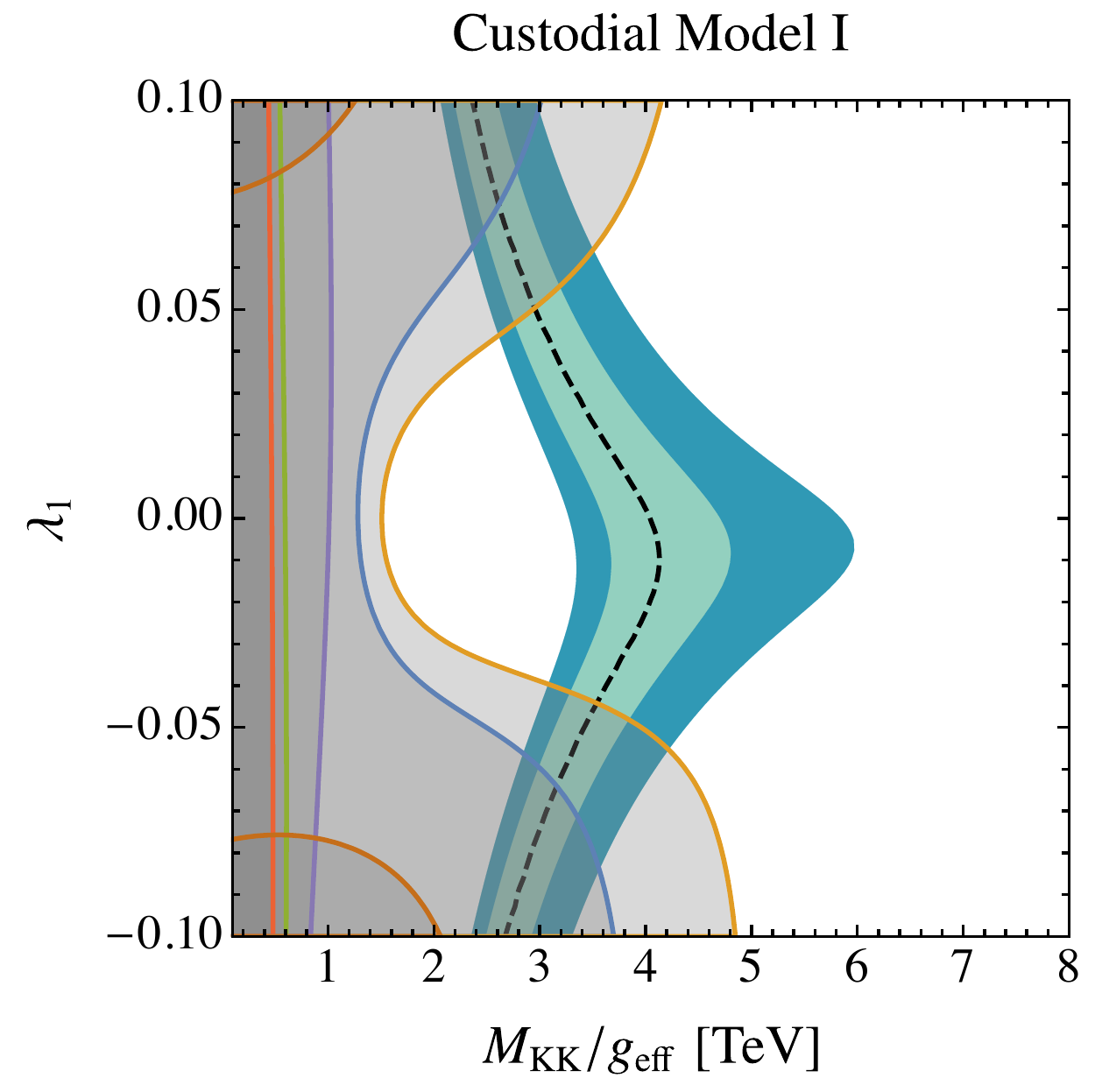} 
\includegraphics[scale=.65]{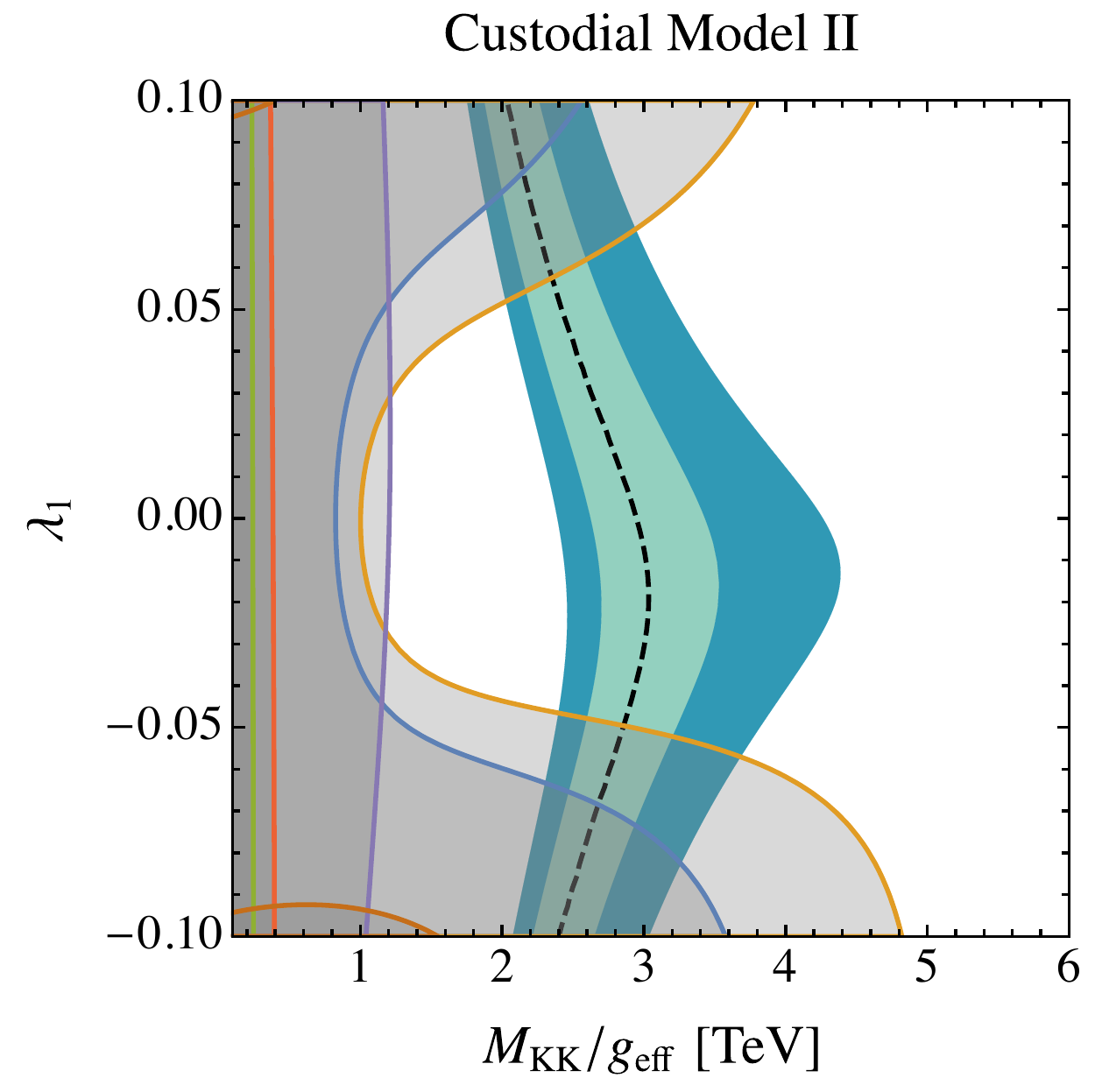} 
\vspace{-2mm}
\caption{\label{fig:portalplots} 
Impact of a Higgs portal coupling $\lambda_1$ on the fit results in the custodial RS models. We use $g_t/g_\text{eff}=-0.5$ for the custodial model~I and $g_t/g_\text{eff}=0$ for the custodial model~II as in (\ref{benchmark}), along with $\beta=1$ and $c_t=0.4$. The constraints from Run~1 resonance searches refer to $WW$ (blue), $ZZ$ (orange), $Z\gamma$ (green), $t\bar t$ (purple), dijets (red) and $hh$ (brown). See text for further explanations.}
\vspace{-2mm}
\end{center}
\end{figure}

The $S\to hh$ decay channel and the admixture of a tree-level coupling to $WW$ and $ZZ$ induced by the mixing with the Higgs boson can have a significant impact on our phenomenological analysis. In Figure~\ref{fig:portalplots}, we present the $1\sigma$ and $2\sigma$ fit regions to the diphoton excess in the custodial RS models in the presence of the portal coupling $\lambda_1$. These plots are analogous to those shown in Figure~\ref{fig:plots}, except that we have fixed the values of $g_t/g_{\rm eff}$ to those of the benchmark points in (\ref{benchmark}). The meaning of the colors of the various curves is the same as before. In all cases the LHC Run~1 bound on the $S\to ZZ$ rate provides the strongest constraint, excluding portal couplings $|\lambda_1|\gtrsim 0.06$~(0.07) in the custodial RS model~I (II). The impact of a small portal coupling $\lambda_1=0.02$ or 0.04 on the various branching ratios is shown in the central and lower portions of Table~\ref{tab:BRs}. Even for such a small coupling the $S\to WW$, $ZZ$ branching ratios can be significantly enhanced, and a sizable $S\to hh$ branching fraction can open up.

\section{\boldmath Three-body decay $S\to t\bar t h$} 
\label{sec:Stth}

\begin{figure}[t]
\begin{center}
\includegraphics[scale=0.82]{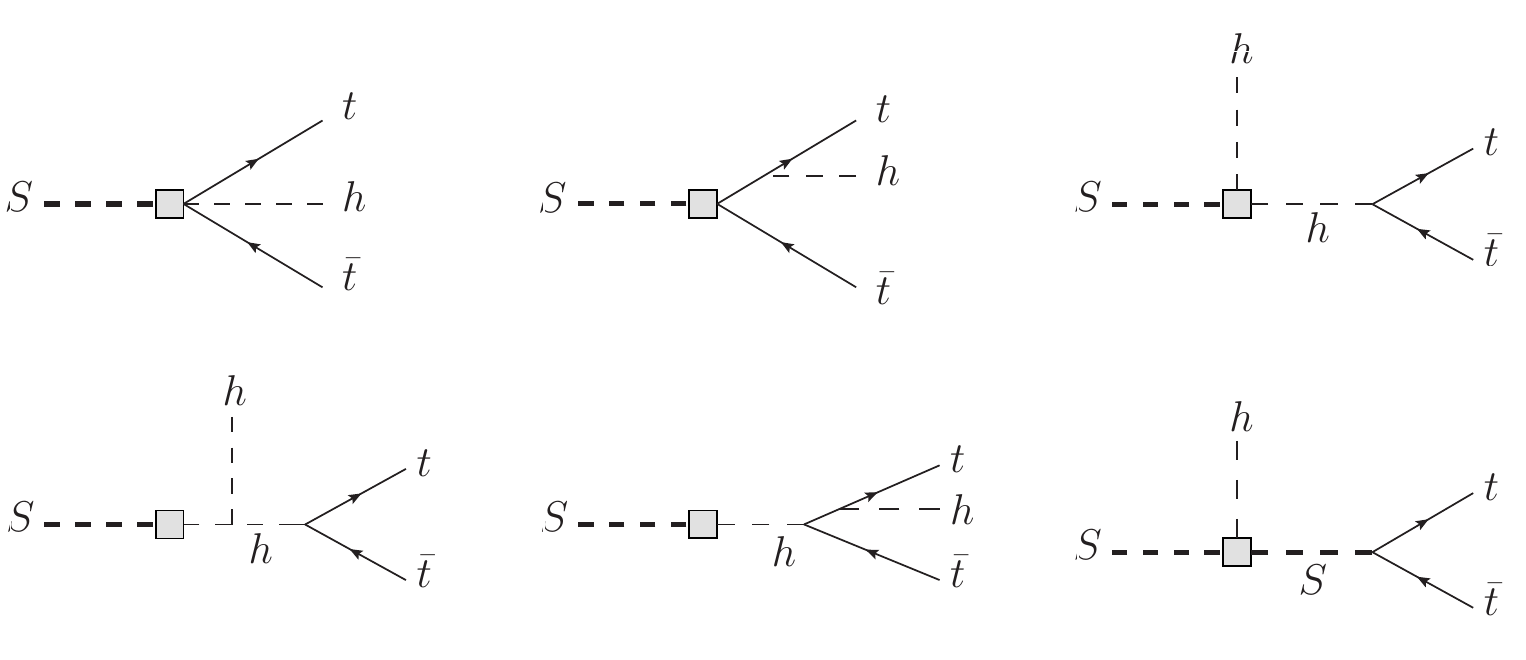}
\caption{\label{fig:tth} 
Diagrams contributing to the decay $S\to t\bar t h$.}
\end{center}
\end{figure}

The effective Lagrangian (\ref{Ltth}) contains a tree-level coupling of the new resonance $S$ to a $t\bar t$ pair and a Higgs boson. It is interesting to ask if this coupling might explain the enhanced $t\bar th$ production rates reported by ATLAS and CMS \cite{CMS:2015kwa}. At tree level, the three-body decay $S\to t\bar t h$ is mediated by the diagrams shown in Figure~\ref{fig:tth}. Note that both portal couplings introduced in (\ref{Leffportals}) contribute here. Introducing the dimensionless variables $z=m_{t\bar t}^2/m_S^2$ and $w=m_{th}^2/m_S^2$, we obtain the Dalitz distribution
\begin{equation}
\begin{aligned}
   \frac{d^2\Gamma(S\to t\bar t h)}{dw\,dz}
   &= \frac{3y_t^2 m_S^3}{256\pi^3}\,c_{tt}^2\,\Big[ A^2 (z-4x_t)
    - 2AB\,x_t (1-2w-z+2x_t+x_h) \\
   &\hspace{2.35cm}\mbox{}+ B^2 x_t\,\Big( (1-w+x_t)(w-x_t-x_h) - z(w-x_t) \Big) \Big] \,,
\end{aligned}
\end{equation}
where
\begin{equation}\label{eq52}
\begin{aligned}
   A &= 1 + \left( \frac{2x_t}{1-w-z+x_t+x_h} + \frac{2x_t}{w-x_t} \right)
    \left( 1 + \frac{1}{m_S\,c_{tt}}\,\frac{\lambda_1}{1-x_h} \right) \\
   &\quad\mbox{}+ \frac{1}{m_S\,c_{tt}}\,\frac{\lambda_1}{z-x_h} \left( 1 + \frac{3x_h}{1-x_h} \right)
    - \frac{v^2}{m_S^2}\,\frac{\lambda_2}{1-z} \,, \\
   B &= \left( \frac{1}{1-w-z+x_t+x_h} - \frac{1}{w-x_t} \right)
    \left( 1 + \frac{1}{m_S\,c_{tt}}\,\frac{\lambda_1}{1-x_h} \right) .
\end{aligned}
\end{equation}
The phase space for the variables $w$ and $z$ is $w_{\rm min}(z)\le w\le w_{\rm max}(z)$ and $4x_t\le z\le(1-\sqrt{x_h})^2$, where
\begin{equation}
   w_{\rm max/min}(z) = \frac{(1-x_h)^2}{4z} - \frac{1}{4z} \left( \sqrt{z(z-4x_t)}
    \mp \sqrt{(1-z-x_h)^2-4z x_h} \right)^2 .
\end{equation}
Our results for the $S\to t\bar t h$ branching ratio obtained by integrating over the Dalitz plot are shown in the penultimate column in Table~\ref{tab:BRs}. This branching ratio is typically two orders of magnitude smaller than the $S\to t\bar t$ branching fraction. This will be true in any model in which the decays of the new resonance can be described in terms of local operators. Given the existing upper bound from $t\bar t$ resonance searches in Run~1 shown in Table~\ref{tab:bounds}, we would expect that the rate for $pp\to S\to t\bar t h$ at $\sqrt{s}=8$~TeV cannot be larger than about 12~fb. This falls short by far to explain the enhanced Higgs production rate in association with $t\bar t$. In our specific models, the predicted $t\bar t h$ production rates do not exceed 8~fb.

\section{Conclusions}
\label{sec:concl}

The recent hint of a 750~GeV diphoton excess, observed in the data from Run~2 of the LHC by both ATLAS and CMS, could be the first direct manifestation of physics beyond the SM. If verified by future analyses, this excess will most likely  have been created by the decay of a new, scalar boson with a mass of 750~GeV produced either in gluon fusion or in $b\bar b$-initiated scattering. Many possible interpretations of such a boson have been proposed. Remarkably, the simple addition of a single new scalar to the SM as well as several well motivated UV completions -- including the Minimal Supersymmetric SM -- have already been excluded. 

In this paper, we have shown that the diphoton signal can be reproduced from a very straightforward extension of the popular warped extra-dimension models introduced by Randall and Sundrum. In these models, our spacetime is enlarged by a warped extra dimension in order to give natural explanations of the gauge and flavor hierarchy problems of the SM. We have identified the diphoton resonance with the lightest excitation of an additional bulk scalar field. Such a scalar might serve as the localizer field providing a dynamical generation of the bulk mass parameters of the 5D fermions and is thus a natural ingredient of RS models with bulk matter fields. In addition to a model with a minimal particle content, we have discussed two implementations of RS models providing a custodial protection mechanism for electroweak precision observables. These models feature a larger number of fundamental 5D fermions compared with the SM. As a result, the gluon-fusion production process and the decay into two photons are enhanced by the large multiplicity of vector-like KK fermion states propagating in the loop. By summing up the contribution from the infinite towers of KK fermions into 5D propagator functions we were able to derive remarkably simple analytic expressions for the effective couplings to gluons or photons, which to very good approximation simply count the number of fermionic degrees of freedom, weighted by group-theory factors. For the custodial RS models, we have found that with ${\cal O}(1)$ couplings of the resonance $S$ to fermions and KK masses in the multi-TeV range one can explain the diphoton signal without violating any of the Run~1 bounds from resonance searches in various diboson and dijet channels. 

Useful side products of our analysis, which can be adapted to any model for the diphoton resonance that at the scale $\mu=m_S$ can be mapped onto an effective Lagrangian with local interactions, are the calculation of the gluon-fusion production cross section $\sigma(pp\to S)$ at NNLO in QCD, an expression for the inclusive $S\to gg$ decay rate at N$^3$LO, a study of the $S\to t\bar t h$ three-body decay mode, and a phenomenological analysis of portal couplings connecting $S$ with the Higgs field.

We conclude that our simple extension of the established RS models can deliver one of the most elegant and minimal explanations for the observed diphoton excess. Assuming that the resonance will survive future verifications, it could hint at the existence of a warped extra dimension and open the door to detailed studies of the parameter space of RS models.

\subsubsection*{Acknowledgements}

We are grateful to Joachim Kopp for useful discussions which have initiated this project, to Daniel Wilhelm for adapting the {\tt CuTe} program to the calculation of the $pp\to S$ cross section, and to Ulrich Haisch for drawing our attention to the possible impact of interference effects in $S\to t\bar t$ resonance searches. M.B.\ acknowledges the support of the Alexander von Humboldt Foundation. The work of C.H.\ and M.N.\ was supported by the Advanced Grant EFT4LHC of the European Research Council (ERC), the DFG Cluster of Excellence {\em Precision Physics, Fundamental Interactions and Structure of Matter} (PRISMAâ EXC 1098), grant 05H12UME of the German Federal Ministry for Education and Research (BMBF) and the DFG Graduate School {\em Symmetry Breaking in Fundamental Interactions} (GRK 1581).

\begin{appendix}

\section{\boldmath RG evolution of the Wilson coefficients $c_{gg}$ and $c_{tt}$}
\label{app:RGmixing}
\renewcommand{\theequation}{A.\arabic{equation}}
\setcounter{equation}{0}

The mixing among the Wilson coefficients $c_{tt}$ and $c_{gg}$ in the effective Lagrangians (\ref{Leff}) and (\ref{Ltth}) is described by the RG equations
\begin{equation}
   \mu\,\frac{d}{d\mu}\,c_{gg}(\mu) = \gamma_{gg}(\alpha_s)\,c_{gg}(\mu) \,, \qquad
   \mu\,\frac{d}{d\mu}\,c_{tt}(\mu) = - \gamma_{tg}(\alpha_s)\,c_{gg}(\mu) \,,
\end{equation}
where $\alpha_s=\alpha_s(\mu)$, and $\gamma_{gg}$ is given by the exact expression \cite{Inami:1982xt,Grinstein:1988wz}  
\begin{equation}
   \gamma_{gg}(\alpha_s) 
   = \alpha_s^2\,\frac{d}{d\alpha_s} \bigg( \frac{\beta(\alpha_s)}{\alpha_s^2} \bigg)
\end{equation}
in terms of the $\beta$-function $\beta(\alpha_s)=\mu\,d\alpha_s(\mu)/d\mu$. The leading contribution to $\gamma_{tg}(\alpha_s)$ has been obtained first in \cite{Collins:1976yq}. The exact solutions to the evolution equations are
\begin{equation}
\begin{aligned}
   c_{gg}(\mu) 
   &= \frac{\beta(\alpha_s(\mu))/\alpha_s^2(\mu)}{\beta(\alpha_s(\mu_0))/\alpha_s^2(\mu_0)}\,
    c_{gg}(\mu_0) \,, \\
   c_{tt}(\mu) 
   &= c_{tt}(\mu_0) 
    - \frac{\alpha_s^2(\mu_0)}{\beta(\alpha_s(\mu_0))}\,c_{gg}(\mu_0)
    \int\limits_{\alpha_s(\mu_0)}^{\alpha_s(\mu)}\!d\alpha\,\frac{\gamma_{tg}(\alpha)}{\alpha^2} \,. 
\end{aligned}
\end{equation}
The perturbative expansions of the anomalous dimension and $\beta$-function read
\begin{equation}
   \frac{\beta(\alpha_s)}{\alpha_s^2} 
    = - \frac{1}{2\pi} \left( \beta_0 + \beta_1\,\frac{\alpha_s}{4\pi} + \dots \right) , \qquad
   \gamma_{tg}(\alpha_s) = \gamma_1 \left( \frac{\alpha_s}{4\pi} \right)^2 + \dots \,,
\end{equation}
where $\gamma_1=24 C_F$ \cite{Collins:1976yq}.

\section{\boldmath Inclusive $S\to gg$ decay rate at N$^3$LO in QCD}
\label{app:Sgg}
\renewcommand{\theequation}{B.\arabic{equation}}
\setcounter{equation}{0}

An analytic expression for the inclusive $S\to gg$ decay rate at N$^3$LO in QCD perturbation theory can be derived using existing calculations of the Higgs-boson decay rate $\Gamma(h\to gg)$ up to ${\cal O}(\alpha_s^5)$ obtained in \cite{Baikov:2006ch,Schreck:2007um}. Taking the heavy top-quark limit and dividing the result by the three-loop expression for the matching coefficient $C_t(m_t,\mu)$ obtained in \cite{Schroder:2005hy,Chetyrkin:2005ia}, we find
\begin{equation}
   \Gamma(S\to gg) = \Gamma_0(\mu) \left[ 1
    + \sum_{n\ge 1} \left( \frac{\alpha_s(\mu)}{\pi} \right)^n g_n(\mu) \right] ,
\end{equation}
with 
\begin{equation}
   \Gamma_0(\mu) = \frac{m_S^3}{8\pi^3}\,\alpha_s^2(\mu)\,c_{gg}^2(\mu) \,,
\end{equation}
and (for $N_c=3$ colors and $n_f=6$ light flavors)
\begin{equation}
\begin{aligned}
   g_1(\mu) &= \frac{45}{4} + \frac{7}{2}\,L_h \,, \\
   g_2(\mu) &= \frac{4313}{32} - \frac{49\pi^2}{16} - \frac{435\zeta_3}{8}
    + \frac{147}{16}\,L_h^2 + \frac{1049}{16}\,L_h \,, \\
   g_3(\mu) &= \frac{2821025}{1728} - \frac{3899\pi^2}{48}
    - \frac{23769\zeta_3}{16} + \frac{10015\zeta_5}{24} \\
   &\quad\mbox{}+ \frac{343}{16}\,L_h^3 + \frac{3899}{16}\,L_h^2 
    + \left( \frac{8279}{8} - \frac{343\pi^2}{16} - \frac{3045\zeta_3}{8} \right) L_h \,,
\end{aligned}
\end{equation}
where $L_h=\ln(\mu^2/m_S^2)$. Numerically, we find for the $K$-factor at $\mu=m_S$ the perturbative expansion $K_{gg}^{\rm N^3LO}=1+0.32768+0.03325-0.01290\approx 1.348$, which exhibits excellent convergence. 

\end{appendix}


\begin{thebibliography}{99}

\bibitem{ATLAS2015-81} 
  The ATLAS collaboration,
  %``Search for resonances decaying to photon pairs in 3.2 fb$^{-1}$ of $pp$ collisions at $\sqrt{s}$ = 13 TeV with the ATLAS detector,''
  ATLAS-CONF-2015-081.
  %%CITATION = ATLAS-CONF-2015-081;%%

\bibitem{CMS:2015dxe} 
  CMS Collaboration [CMS Collaboration],
  %``Search for new physics in high mass diphoton events in proton-proton
  %collisions at 13TeV,''
  CMS-PAS-EXO-15-004.
  %%CITATION = CMS-PAS-EXO-15-004;%%

\bibitem{Buttazzo:2015txu} 
  D.~Buttazzo, A.~Greljo and D.~Marzocca,
  %``Knocking on new physicsÕ door with a scalar resonance,''
  Eur.\ Phys.\ J.\ C {\bf 76}, no. 3, 116 (2016)
  doi:10.1140/epjc/s10052-016-3970-7
  [arXiv:1512.04929 [hep-ph]].
  %%CITATION = doi:10.1140/epjc/s10052-016-3970-7;%%

\bibitem{Staub:2016dxq} 
  F.~Staub {\it et al.},
  %``Precision tools and models to narrow in on the 750 GeV diphoton resonance,''
  arXiv:1602.05581 [hep-ph].
  %%CITATION = ARXIV:1602.05581;%%
  
\bibitem{Franceschini:2015kwy} 
  R.~Franceschini {\it et al.},
  %``What is the $\gamma \gamma$ resonance at 750 GeV?,''
  JHEP {\bf 1603}, 144 (2016)
  doi:10.1007/JHEP03(2016)144
  [arXiv:1512.04933 [hep-ph]].
  %%CITATION = doi:10.1007/JHEP03(2016)144;%%
  
\bibitem{Fichet:2015vvy} 
  S.~Fichet, G.~von Gersdorff and C.~Royon,
  %``Scattering light by light at 750 GeV at the LHC,''
  Phys.\ Rev.\ D {\bf 93}, no. 7, 075031 (2016)
  doi:10.1103/PhysRevD.93.075031
  [arXiv:1512.05751 [hep-ph]].
  %%CITATION = doi:10.1103/PhysRevD.93.075031;%%
  
\bibitem{Csaki:2015vek} 
  C.~Cs\'aki, J.~Hubisz and J.~Terning,
  %``Minimal model of a diphoton resonance: Production without gluon couplings,''
  Phys.\ Rev.\ D {\bf 93}, no. 3, 035002 (2016)
  doi:10.1103/PhysRevD.93.035002
  [arXiv:1512.05776 [hep-ph]].
  %%CITATION = doi:10.1103/PhysRevD.93.035002;%%

\bibitem{Gupta:2015zzs} 
  R.~S.~Gupta, S.~J\"ager, Y.~Kats, G.~Perez and E.~Stamou,
  %``Interpreting a 750 GeV Diphoton Resonance,''
  arXiv:1512.05332 [hep-ph].
  %%CITATION = ARXIV:1512.05332;%%

\bibitem{Petersson:2015mkr} 
  C.~Petersson and R.~Torre,
  %``750 GeV Diphoton Excess from the Goldstino Superpartner,''
  Phys.\ Rev.\ Lett.\  {\bf 116}, no. 15, 151804 (2016)
  doi:10.1103/PhysRevLett.116.151804
  [arXiv:1512.05333 [hep-ph]].
  %%CITATION = doi:10.1103/PhysRevLett.116.151804;%%
  
\bibitem{Ding:2015rxx} 
  R.~Ding, L.~Huang, T.~Li and B.~Zhu,
  %``Interpreting $750$ GeV Diphoton Excess with R-parity Violation Supersymmetry,''
  arXiv:1512.06560 [hep-ph].
  %%CITATION = ARXIV:1512.06560;%%

\bibitem{Allanach:2015ixl} 
  B.~C.~Allanach, P.~S.~B.~Dev, S.~A.~Renner and K.~Sakurai,
  %``750 GeV diphoton excess explained by a resonant sneutrino in R-parity violating supersymmetry,''
  Phys.\ Rev.\ D {\bf 93}, no. 11, 115022 (2016)
  doi:10.1103/PhysRevD.93.115022
  [arXiv:1512.07645 [hep-ph]].
  %%CITATION = doi:10.1103/PhysRevD.93.115022;%%
  
\bibitem{Harigaya:2015ezk} 
  K.~Harigaya and Y.~Nomura,
  %``Composite Models for the 750 GeV Diphoton Excess,''
  Phys.\ Lett.\ B {\bf 754}, 151 (2016)
  doi:10.1016/j.physletb.2016.01.026
  [arXiv:1512.04850 [hep-ph]].
  %%CITATION = doi:10.1016/j.physletb.2016.01.026;%%
  
\bibitem{Belyaev:2015hgo} 
  A.~Belyaev, G.~Cacciapaglia, H.~Cai, T.~Flacke, A.~Parolini and H.~Serodio,
  %``Singlets in Composite Higgs Models in light of the LHC di-photon searches,''
  arXiv:1512.07242 [hep-ph].
  %%CITATION = ARXIV:1512.07242;%%

\bibitem{Son:2015vfl} 
  M.~Son and A.~Urbano,
  %``A new scalar resonance at 750 GeV: Towards a proof of concept in favor of strongly interacting theories,''
  JHEP {\bf 1605}, 181 (2016)
  doi:10.1007/JHEP05(2016)181
  [arXiv:1512.08307 [hep-ph]].
  %%CITATION = doi:10.1007/JHEP05(2016)181;%%

\bibitem{Harigaya:2016pnu} 
  K.~Harigaya and Y.~Nomura,
  %``A Composite Model for the 750 GeV Diphoton Excess,''
  JHEP {\bf 1603}, 091 (2016)
  doi:10.1007/JHEP03(2016)091
  [arXiv:1602.01092 [hep-ph]].
  %%CITATION = doi:10.1007/JHEP03(2016)091;%%

\bibitem{Cox:2015ckc} 
  P.~Cox, A.~D.~Medina, T.~S.~Ray and A.~Spray,
  %``Diphoton Excess at 750 GeV from a Radion in the Bulk-Higgs Scenario,''
  arXiv:1512.05618 [hep-ph].
  %%CITATION = ARXIV:1512.05618;%%
  
\bibitem{Ahmed:2015uqt} 
  A.~Ahmed, B.~M.~Dillon, B.~Grzadkowski, J.~F.~Gunion and Y.~Jiang,
  %``Higgs-radion interpretation of 750 GeV di-photon excess at the LHC,''
  arXiv:1512.05771 [hep-ph].
  %%CITATION = ARXIV:1512.05771;%%

\bibitem{Bardhan:2015hcr} 
  D.~Bardhan, D.~Bhatia, A.~Chakraborty, U.~Maitra, S.~Raychaudhuri and T.~Samui,
  %``Radion Candidate for the LHC Diphoton Resonance,''
  arXiv:1512.06674 [hep-ph].
  %%CITATION = ARXIV:1512.06674;%%

\bibitem{Aloni:2015mxa} 
  D.~Aloni, K.~Blum, A.~Dery, A.~Efrati and Y.~Nir,
  %``On a possible large width 750 GeV diphoton resonance at ATLAS and CMS,''
  arXiv:1512.05778 [hep-ph].
  %%CITATION = ARXIV:1512.05778;%%

\bibitem{Bauer:2015boy} 
  M.~Bauer and M.~Neubert,
  %``Flavor anomalies, the 750 GeV diphoton excess, and a dark matter candidate,''
  Phys.\ Rev.\ D {\bf 93}, no. 11, 115030 (2016)
  doi:10.1103/PhysRevD.93.115030
  [arXiv:1512.06828 [hep-ph]].
  %%CITATION = doi:10.1103/PhysRevD.93.115030;%%
  
\bibitem{Murphy:2015kag} 
  C.~W.~Murphy,
  %``Vector Leptoquarks and the 750 GeV Diphoton Resonance at the LHC,''
  Phys.\ Lett.\ B {\bf 757}, 192 (2016)
  doi:10.1016/j.physletb.2016.03.076
  [arXiv:1512.06976 [hep-ph]].
  %%CITATION = doi:10.1016/j.physletb.2016.03.076;%%
  
\bibitem{Goertz:2015nkp} 
  F.~Goertz, J.~F.~Kamenik, A.~Katz and M.~Nardecchia,
  %``Indirect Constraints on the Scalar Di-Photon Resonance at the LHC,''
  JHEP {\bf 1605}, 187 (2016)
  doi:10.1007/JHEP05(2016)187
  [arXiv:1512.08500 [hep-ph]].
  %%CITATION = doi:10.1007/JHEP05(2016)187;%%
  
\bibitem{Dev:2015vjd} 
  P.~S.~B.~Dev, R.~N.~Mohapatra and Y.~Zhang,
  %``Quark Seesaw, Vectorlike Fermions and Diphoton Excess,''
  JHEP {\bf 1602}, 186 (2016)
  doi:10.1007/JHEP02(2016)186
  [arXiv:1512.08507 [hep-ph]].
  %%CITATION = doi:10.1007/JHEP02(2016)186;%%
  
\bibitem{Hernandez:2015hrt} 
  A.~E.~C.~Hern\'andez,
  %``The 750 GeV diphoton resonance can cause the SM fermion mass and mixing pattern,''
  arXiv:1512.09092 [hep-ph].
  %%CITATION = ARXIV:1512.09092;%%

\bibitem{Belanger:2016ywb} 
  G.~BŽlanger and C.~Delaunay,
  %``A Dark Sector for $g_\mu-2$, $R_K$ and a Diphoton Resonance,''
  arXiv:1603.03333 [hep-ph].
  %%CITATION = ARXIV:1603.03333;%%  

\bibitem{Randall:1999ee} 
  L.~Randall and R.~Sundrum,
  %``A Large mass hierarchy from a small extra dimension,''
  Phys.\ Rev.\ Lett.\  {\bf 83}, 3370 (1999)
  doi:10.1103/PhysRevLett.83.3370
  [hep-ph/9905221].
  %%CITATION = doi:10.1103/PhysRevLett.83.3370;%%

\bibitem{Grossman:1999ra} 
  Y.~Grossman and M.~Neubert,
  %``Neutrino masses and mixings in nonfactorizable geometry,''
  Phys.\ Lett.\ B {\bf 474}, 361 (2000)
  doi:10.1016/S0370-2693(00)00054-X
  [hep-ph/9912408].
  %%CITATION = doi:10.1016/S0370-2693(00)00054-X;%%
  
\bibitem{Gherghetta:2000qt} 
  T.~Gherghetta and A.~Pomarol,
  %``Bulk fields and supersymmetry in a slice of AdS,''
  Nucl.\ Phys.\ B {\bf 586}, 141 (2000)
  doi:10.1016/S0550-3213(00)00392-8
  [hep-ph/0003129].
  %%CITATION = doi:10.1016/S0550-3213(00)00392-8;%%

\bibitem{Kaplan:2001ga} 
  D.~E.~Kaplan and T.~M.~P.~Tait,
  %``New tools for fermion masses from extra dimensions,''
  JHEP {\bf 0111}, 051 (2001)
  doi:10.1088/1126-6708/2001/11/051
  [hep-ph/0110126].
  %%CITATION = doi:10.1088/1126-6708/2001/11/051;%%

\bibitem{Cai:2015hzc} 
  C.~Cai, Z.~H.~Yu and H.~H.~Zhang,
  %``750 GeV diphoton resonance as a singlet scalar in an extra dimensional model,''
  Phys.\ Rev.\ D {\bf 93}, no. 7, 075033 (2016)
  doi:10.1103/PhysRevD.93.075033
  [arXiv:1512.08440 [hep-ph]].
  %%CITATION = doi:10.1103/PhysRevD.93.075033;%%

\bibitem{Abel:2016pyc} 
  S.~Abel and V.~V.~Khoze,
  %``Photo-production of a 750 GeV di-photon resonance mediated by Kaluza-Klein leptons in the loop,''
  JHEP {\bf 1605}, 063 (2016)
  doi:10.1007/JHEP05(2016)063
  [arXiv:1601.07167 [hep-ph]].
  %%CITATION = doi:10.1007/JHEP05(2016)063;%%
    
\bibitem{Arun:2015ubr} 
  M.~T.~Arun and P.~Saha,
  %``Gravitons in multiply warped scenarios - at 750 GeV and beyond,''
  arXiv:1512.06335 [hep-ph].
  %%CITATION = ARXIV:1512.06335;%%
  
\bibitem{Geng:2016xin} 
  C.~Q.~Geng and D.~Huang,
  %``Note on spin-2 particle interpretation of the 750 GeV diphoton excess,''
  Phys.\ Rev.\ D {\bf 93}, no. 11, 115032 (2016)
  doi:10.1103/PhysRevD.93.115032
  [arXiv:1601.07385 [hep-ph]].
  %%CITATION = doi:10.1103/PhysRevD.93.115032;%%
  
\bibitem{Giddings:2016sfr} 
  S.~B.~Giddings and H.~Zhang,
  %``Kaluza-Klein graviton phenomenology for warped compactifications, and the 750 GeV diphoton excess,''
  Phys.\ Rev.\ D {\bf 93}, no. 11, 115002 (2016)
  doi:10.1103/PhysRevD.93.115002
  [arXiv:1602.02793 [hep-ph]].
  %%CITATION = doi:10.1103/PhysRevD.93.115002;%%
  
\bibitem{Goldberger:1999uk}
  W.~D.~Goldberger and M.~B.~Wise,
  %``Modulus stabilization with bulk fields,''
  Phys.\ Rev.\ Lett.\  {\bf 83}, 4922 (1999)
  doi:10.1103/PhysRevLett.83.4922
  [hep-ph/9907447].
  %%CITATION = doi:10.1103/PhysRevLett.83.4922;%%

\bibitem{Huber:2000ie} 
  S.~J.~Huber and Q.~Shafi,
  %``Fermion masses, mixings and proton decay in a Randall-Sundrum model,''
  Phys.\ Lett.\ B {\bf 498}, 256 (2001)
  doi:10.1016/S0370-2693(00)01399-X
  [hep-ph/0010195].
  %%CITATION = doi:10.1016/S0370-2693(00)01399-X;%%
    
\bibitem{Csaki:2002gy} 
  C.~Csaki, J.~Erlich and J.~Terning,
  %``The Effective Lagrangian in the Randall-Sundrum model and electroweak physics,''
  Phys.\ Rev.\ D {\bf 66}, 064021 (2002)
  doi:10.1103/PhysRevD.66.064021
  [hep-ph/0203034].
  %%CITATION = doi:10.1103/PhysRevD.66.064021;%%

\bibitem{Carena:2003fx} 
  M.~Carena, A.~Delgado, E.~Ponton, T.~M.~P.~Tait and C.~E.~M.~Wagner,
  %``Precision electroweak data and unification of couplings in warped extra dimensions,''
  Phys.\ Rev.\ D {\bf 68}, 035010 (2003)
  doi:10.1103/PhysRevD.68.035010
  [hep-ph/0305188].
  %%CITATION = doi:10.1103/PhysRevD.68.035010;%%
  
\bibitem{Malm:2013jia} 
  R.~Malm, M.~Neubert, K.~Novotny and C.~Schmell,
  %``5D Perspective on Higgs Production at the Boundary of a Warped Extra Dimension,''
  JHEP {\bf 1401}, 173 (2014)
  doi:10.1007/JHEP01(2014)173
  [arXiv:1303.5702 [hep-ph]].
  %%CITATION = doi:10.1007/JHEP01(2014)173;%%

\bibitem{Davoudiasl:1999tf} 
  H.~Davoudiasl, J.~L.~Hewett and T.~G.~Rizzo,
  %``Bulk gauge fields in the Randall-Sundrum model,''
  Phys.\ Lett.\ B {\bf 473}, 43 (2000)
  doi:10.1016/S0370-2693(99)01430-6
  [hep-ph/9911262].
  %%CITATION = doi:10.1016/S0370-2693(99)01430-6;%%
  
\bibitem{Agashe:2003zs} 
  K.~Agashe, A.~Delgado, M.~J.~May and R.~Sundrum,
  %``RS1, custodial isospin and precision tests,''
  JHEP {\bf 0308}, 050 (2003)
  doi:10.1088/1126-6708/2003/08/050
  [hep-ph/0308036].
  %%CITATION = doi:10.1088/1126-6708/2003/08/050;%%
  
\bibitem{Csaki:2003zu} 
  C.~Csaki, C.~Grojean, L.~Pilo and J.~Terning,
  %``Towards a realistic model of Higgsless electroweak symmetry breaking,''
  Phys.\ Rev.\ Lett.\  {\bf 92}, 101802 (2004)
  doi:10.1103/PhysRevLett.92.101802
  [hep-ph/0308038].
  %%CITATION = doi:10.1103/PhysRevLett.92.101802;%%

\bibitem{Agashe:2006at} 
  K.~Agashe, R.~Contino, L.~Da Rold and A.~Pomarol,
  %``A Custodial symmetry for $Zb \bar b$,''
  Phys.\ Lett.\ B {\bf 641}, 62 (2006)
  doi:10.1016/j.physletb.2006.08.005
  [hep-ph/0605341].
  %%CITATION = doi:10.1016/j.physletb.2006.08.005;%%

\bibitem{Blanke:2008zb} 
  M.~Blanke, A.~J.~Buras, B.~Duling, S.~Gori and A.~Weiler,
  %``$\Delta$ F=2 Observables and Fine-Tuning in a Warped Extra Dimension with Custodial Protection,''
  JHEP {\bf 0903}, 001 (2009)
  doi:10.1088/1126-6708/2009/03/001
  [arXiv:0809.1073 [hep-ph]].
  %%CITATION = doi:10.1088/1126-6708/2009/03/001;%%
  
\bibitem{Albrecht:2009xr} 
  M.~E.~Albrecht, M.~Blanke, A.~J.~Buras, B.~Duling and K.~Gemmler,
  %``Electroweak and Flavour Structure of a Warped Extra Dimension with Custodial Protection,''
  JHEP {\bf 0909}, 064 (2009)
  doi:10.1088/1126-6708/2009/09/064
  [arXiv:0903.2415 [hep-ph]].
  %%CITATION = doi:10.1088/1126-6708/2009/09/064;%%

\bibitem{Casagrande:2010si} 
  S.~Casagrande, F.~Goertz, U.~Haisch, M.~Neubert and T.~Pfoh,
  %``The Custodial Randall-Sundrum Model: From Precision Tests to Higgs Physics,''
  JHEP {\bf 1009}, 014 (2010)
  doi:10.1007/JHEP09(2010)014
  [arXiv:1005.4315 [hep-ph]].
  %%CITATION = doi:10.1007/JHEP09(2010)014;%%
  
\bibitem{Hahn:2013nza} 
  J.~Hahn, C.~H\"orner, R.~Malm, M.~Neubert, K.~Novotny and C.~Schmell,
  %``Higgs Decay into Two Photons at the Boundary of a Warped Extra Dimension,''
  Eur.\ Phys.\ J.\ C {\bf 74}, no. 5, 2857 (2014)
  doi:10.1140/epjc/s10052-014-2857-8
  [arXiv:1312.5731 [hep-ph]].
  %%CITATION = doi:10.1140/epjc/s10052-014-2857-8;%%

\bibitem{Agashe:2004cp} 
  K.~Agashe, G.~Perez and A.~Soni,
  %``Flavor structure of warped extra dimension models,''
  Phys.\ Rev.\ D {\bf 71}, 016002 (2005)
  doi:10.1103/PhysRevD.71.016002
  [hep-ph/0408134].
  %%CITATION = doi:10.1103/PhysRevD.71.016002;%%
  
\bibitem{Csaki:2008zd} 
  C.~Csaki, A.~Falkowski and A.~Weiler,
  %``The Flavor of the Composite Pseudo-Goldstone Higgs,''
  JHEP {\bf 0809}, 008 (2008)
  doi:10.1088/1126-6708/2008/09/008
  [arXiv:0804.1954 [hep-ph]].
  %%CITATION = doi:10.1088/1126-6708/2008/09/008;%%

\bibitem{Bauer:2009cf} 
  M.~Bauer, S.~Casagrande, U.~Haisch and M.~Neubert,
  %``Flavor Physics in the Randall-Sundrum Model: II. Tree-Level Weak-Interaction Processes,''
  JHEP {\bf 1009}, 017 (2010)
  doi:10.1007/JHEP09(2010)017
  [arXiv:0912.1625 [hep-ph]].
  %%CITATION = doi:10.1007/JHEP09(2010)017;%%

\bibitem{Azatov:2010pf} 
  A.~Azatov, M.~Toharia and L.~Zhu,
  %``Higgs Production from Gluon Fusion in Warped Extra Dimensions,''
  Phys.\ Rev.\ D {\bf 82}, 056004 (2010)
  doi:10.1103/PhysRevD.82.056004
  [arXiv:1006.5939 [hep-ph]].
  %%CITATION = doi:10.1103/PhysRevD.82.056004;%%

\bibitem{Archer:2014jca} 
  P.~R.~Archer, M.~Carena, A.~Carmona and M.~Neubert,
  %``Higgs Production and Decay in Models of a Warped Extra Dimension with a Bulk Higgs,''
  JHEP {\bf 1501}, 060 (2015)
  doi:10.1007/JHEP01(2015)060
  [arXiv:1408.5406 [hep-ph]].
  %%CITATION = doi:10.1007/JHEP01(2015)060;%%

\bibitem{Malm:2014gha} 
  R.~Malm, M.~Neubert and C.~Schmell,
  %``Higgs Couplings and Phenomenology in a Warped Extra Dimension,''
  JHEP {\bf 1502}, 008 (2015)
  doi:10.1007/JHEP02(2015)008
  [arXiv:1408.4456 [hep-ph]].
  %%CITATION = doi:10.1007/JHEP02(2015)008;%%
  
\bibitem{Bauer:2011ah} 
  M.~Bauer, R.~Malm and M.~Neubert,
  %``A Solution to the Flavor Problem of Warped Extra-Dimension Models,''
  Phys.\ Rev.\ Lett.\  {\bf 108}, 081603 (2012)
  doi:10.1103/PhysRevLett.108.081603
  [arXiv:1110.0471 [hep-ph]].
  %%CITATION = doi:10.1103/PhysRevLett.108.081603;%%

\bibitem{Casagrande:2008hr} 
  S.~Casagrande, F.~Goertz, U.~Haisch, M.~Neubert and T.~Pfoh,
  %``Flavor Physics in the Randall-Sundrum Model: I. Theoretical Setup and Electroweak Precision Tests,''
  JHEP {\bf 0810}, 094 (2008)
  doi:10.1088/1126-6708/2008/10/094
  [arXiv:0807.4937 [hep-ph]].
  %%CITATION = doi:10.1088/1126-6708/2008/10/094;%%
  
\bibitem{Cacciapaglia:2006mz} 
  G.~Cacciapaglia, C.~Csaki, G.~Marandella and J.~Terning,
  %``The Gaugephobic Higgs,''
  JHEP {\bf 0702}, 036 (2007)
  doi:10.1088/1126-6708/2007/02/036
  [hep-ph/0611358].
  %%CITATION = doi:10.1088/1126-6708/2007/02/036;%%

\bibitem{Archer:2012qa} 
  P.~R.~Archer,
  %``The Fermion Mass Hierarchy in Models with Warped Extra Dimensions and a Bulk Higgs,''
  JHEP {\bf 1209}, 095 (2012)
  doi:10.1007/JHEP09(2012)095
  [arXiv:1204.4730 [hep-ph]].
  %%CITATION = doi:10.1007/JHEP09(2012)095;%%
  
\bibitem{Breitenlohner:1982jf} 
  P.~Breitenlohner and D.~Z.~Freedman,
  %``Stability in Gauged Extended Supergravity,''
  Annals Phys.\  {\bf 144}, 249 (1982)
  doi:10.1016/0003-4916(82)90116-6.
  %%CITATION = doi:10.1016/0003-4916(82)90116-6;%%

\bibitem{Inami:1982xt} 
  T.~Inami, T.~Kubota and Y.~Okada,
  %``Effective Gauge Theory and the Effect of Heavy Quarks in Higgs Boson Decays,''
  Z.\ Phys.\ C {\bf 18}, 69 (1983)
  doi:10.1007/BF01571710.
  %%CITATION = doi:10.1007/BF01571710;%%\cite{add_refs}

\bibitem{Grinstein:1988wz} 
  B.~Grinstein and L.~Randall,
  %``The Renormalization of $g^{2}$,''
  Phys.\ Lett.\ B {\bf 217}, 335 (1989)
  doi:10.1016/0370-2693(89)90877-0.
  %%CITATION = doi:10.1016/0370-2693(89)90877-0;%%

\bibitem{Collins:1976yq} 
  J.~C.~Collins, A.~Duncan and S.~D.~Joglekar,
  %``Trace and Dilatation Anomalies in Gauge Theories,''
  Phys.\ Rev.\ D {\bf 16}, 438 (1977)
  doi:10.1103/PhysRevD.16.438.
  %%CITATION = doi:10.1103/PhysRevD.16.438;%%

\bibitem{Schroder:2005hy} 
  Y.~Schr\"oder and M.~Steinhauser,
  %``Four-loop decoupling relations for the strong coupling,''
  JHEP {\bf 0601}, 051 (2006)
  doi:10.1088/1126-6708/2006/01/051
  [hep-ph/0512058].
  %%CITATION = doi:10.1088/1126-6708/2006/01/051;%%

\bibitem{Chetyrkin:2005ia} 
  K.~G.~Chetyrkin, J.~H.~K\"uhn and C.~Sturm,
  %``QCD decoupling at four loops,''
  Nucl.\ Phys.\ B {\bf 744}, 121 (2006)
  doi:10.1016/j.nuclphysb.2006.03.020
  [hep-ph/0512060].
  %%CITATION = doi:10.1016/j.nuclphysb.2006.03.020;%%

\bibitem{Azatov:2009na} 
  A.~Azatov, M.~Toharia and L.~Zhu,
  %``Higgs Mediated FCNC's in Warped Extra Dimensions,''
  Phys.\ Rev.\ D {\bf 80}, 035016 (2009)
  doi:10.1103/PhysRevD.80.035016
  [arXiv:0906.1990 [hep-ph]].
  %%CITATION = doi:10.1103/PhysRevD.80.035016;%%

\bibitem{Randall:2001gb} 
  L.~Randall and M.~D.~Schwartz,
  %``Quantum field theory and unification in AdS5,''
  JHEP {\bf 0111}, 003 (2001)
  doi:10.1088/1126-6708/2001/11/003
  [hep-th/0108114].
  %%CITATION = doi:10.1088/1126-6708/2001/11/003;%%
  
\bibitem{Contino:2004vy} 
  R.~Contino and A.~Pomarol,
  %``Holography for fermions,''
  JHEP {\bf 0411}, 058 (2004)
  doi:10.1088/1126-6708/2004/11/058
  [hep-th/0406257].
  %%CITATION = doi:10.1088/1126-6708/2004/11/058;%%
  
\bibitem{Carena:2004zn} 
  M.~Carena, A.~Delgado, E.~Ponton, T.~M.~P.~Tait and C.~E.~M.~Wagner,
  %``Warped fermions and precision tests,''
  Phys.\ Rev.\ D {\bf 71}, 015010 (2005)
  doi:10.1103/PhysRevD.71.015010
  [hep-ph/0410344].
  %%CITATION = doi:10.1103/PhysRevD.71.015010;%%  
  
\bibitem{Csaki:2010aj} 
  C.~Csaki, Y.~Grossman, P.~Tanedo and Y.~Tsai,
  %``Warped penguin diagrams,''
  Phys.\ Rev.\ D {\bf 83}, 073002 (2011)
  doi:10.1103/PhysRevD.83.073002
  [arXiv:1004.2037 [hep-ph]].
  %%CITATION = doi:10.1103/PhysRevD.83.073002;%%  
  
\bibitem{Carena:2012fk} 
  M.~Carena, S.~Casagrande, F.~Goertz, U.~Haisch and M.~Neubert,
  %``Higgs Production in a Warped Extra Dimension,''
  JHEP {\bf 1208}, 156 (2012)
  doi:10.1007/JHEP08(2012)156
  [arXiv:1204.0008 [hep-ph]].
  %%CITATION = doi:10.1007/JHEP08(2012)156;%%
  
\bibitem{CuTe}
  The program {\tt CuTe} and a manual are available for downloaded at:\\
  {\tt https://cute.hepforge.org}

\bibitem{Becher:2011xn} 
  T.~Becher, M.~Neubert and D.~Wilhelm,
  %``Electroweak Gauge-Boson Production at Small $q_T$: Infrared Safety from the Collinear Anomaly,''
  JHEP {\bf 1202}, 124 (2012)
  doi:10.1007/JHEP02(2012)124
  [arXiv:1109.6027 [hep-ph]].
  %%CITATION = doi:10.1007/JHEP02(2012)124;%%
  
\bibitem{Becher:2012yn} 
  T.~Becher, M.~Neubert and D.~Wilhelm,
  %``Higgs-Boson Production at Small Transverse Momentum,''
  JHEP {\bf 1305}, 110 (2013)
  doi:10.1007/JHEP05(2013)110
  [arXiv:1212.2621 [hep-ph]].
  %%CITATION = doi:10.1007/JHEP05(2013)110;%%

\bibitem{Martin:2009iq} 
  A.~D.~Martin, W.~J.~Stirling, R.~S.~Thorne and G.~Watt,
  %``Parton distributions for the LHC,''
  Eur.\ Phys.\ J.\ C {\bf 63}, 189 (2009)
  doi:10.1140/epjc/s10052-009-1072-5
  [arXiv:0901.0002 [hep-ph]].
  %%CITATION = doi:10.1140/epjc/s10052-009-1072-5;%%  

\bibitem{Ball:2014uwa} 
  R.~D.~Ball {\it et al.} [NNPDF Collaboration],
  %``Parton distributions for the LHC Run II,''
  JHEP {\bf 1504}, 040 (2015)
  doi:10.1007/JHEP04(2015)040
  [arXiv:1410.8849 [hep-ph]].
  %%CITATION = doi:10.1007/JHEP04(2015)040;%%

\bibitem{Butterworth:2015oua} 
  J.~Butterworth {\it et al.},
  %``PDF4LHC recommendations for LHC Run II,''
  J.\ Phys.\ G {\bf 43}, 023001 (2016)
  doi:10.1088/0954-3899/43/2/023001
  [arXiv:1510.03865 [hep-ph]].
  %%CITATION = doi:10.1088/0954-3899/43/2/023001;%%

\bibitem{Cooper-Sarkar:2015boa} 
  A.~M.~Cooper-Sarkar,
  %``HERA Collider Results,''
  PoS DIS {\bf 2015}, 005 (2015)
  [arXiv:1507.03849 [hep-ph]].
  %%CITATION = ARXIV:1507.03849;%%
  
\bibitem{Harland-Lang:2014zoa} 
  L.~A.~Harland-Lang, A.~D.~Martin, P.~Motylinski and R.~S.~Thorne,
  %``Parton distributions in the LHC era: MMHT 2014 PDFs,''
  Eur.\ Phys.\ J.\ C {\bf 75}, no. 5, 204 (2015)
  doi:10.1140/epjc/s10052-015-3397-6
  [arXiv:1412.3989 [hep-ph]].
  %%CITATION = doi:10.1140/epjc/s10052-015-3397-6;%%
  
\bibitem{Baikov:2006ch} 
  P.~A.~Baikov and K.~G.~Chetyrkin,
  %``Top Quark Mediated Higgs Boson Decay into Hadrons to Order $\alpha_s^5$,''
  Phys.\ Rev.\ Lett.\  {\bf 97}, 061803 (2006)
  doi:10.1103/PhysRevLett.97.061803
  [hep-ph/0604194].
  %%CITATION = doi:10.1103/PhysRevLett.97.061803;%%

\bibitem{Schreck:2007um} 
  M.~Schreck and M.~Steinhauser,
  %``Higgs Decay to Gluons at NNLO,''
  Phys.\ Lett.\ B {\bf 655}, 148 (2007)
  doi:10.1016/j.physletb.2007.08.080
  [arXiv:0708.0916 [hep-ph]].
  %%CITATION = doi:10.1016/j.physletb.2007.08.080;%%

\bibitem{UliHaisch}
  U.~Haisch, private communication.
  
\bibitem{Dicus:1994bm} 
  D.~Dicus, A.~Stange and S.~Willenbrock,
  %``Higgs decay to top quarks at hadron colliders,''
  Phys.\ Lett.\ B {\bf 333}, 126 (1994)
  doi:10.1016/0370-2693(94)91017-0
  [hep-ph/9404359].
  %%CITATION = doi:10.1016/0370-2693(94)91017-0;%%
    
\bibitem{Craig:2015jba} 
  N.~Craig, F.~D'Eramo, P.~Draper, S.~Thomas and H.~Zhang,
  %``The Hunt for the Rest of the Higgs Bosons,''
  JHEP {\bf 1506}, 137 (2015)
  doi:10.1007/JHEP06(2015)137
  [arXiv:1504.04630 [hep-ph]].
  %%CITATION = doi:10.1007/JHEP06(2015)137;%%

\bibitem{Gori:2016zto} 
  S.~Gori, I.~W.~Kim, N.~R.~Shah and K.~M.~Zurek,
  %``Closing the Wedge: Search Strategies for Extended Higgs Sectors with Heavy Flavor Final States,''
  Phys.\ Rev.\ D {\bf 93}, no. 7, 075038 (2016)
  doi:10.1103/PhysRevD.93.075038
  [arXiv:1602.02782 [hep-ph]].
  %%CITATION = doi:10.1103/PhysRevD.93.075038;%%
    
\bibitem{Buckley:2016mbr} 
  M.~R.~Buckley,
  %``Wide or narrow? The phenomenology of 750 GeV diphotons,''
  Eur.\ Phys.\ J.\ C {\bf 76}, no. 6, 345 (2016)
  doi:10.1140/epjc/s10052-016-4201-y
  [arXiv:1601.04751 [hep-ph]].
  %%CITATION = doi:10.1140/epjc/s10052-016-4201-y;%%
  
\bibitem{CMS:2015neg} 
  CMS Collaboration [CMS Collaboration],
  %``Search for Resonances Decaying to Dijet Final States at $\sqrt{s} = 8$ TeV with Scouting Data,''
  CMS-PAS-EXO-14-005.
  %%CITATION = CMS-PAS-EXO-14-005;%%
  
\bibitem{Aad:2015agg} 
  G.~Aad {\it et al.} [ATLAS Collaboration],
  %``Search for a high-mass Higgs boson decaying to a $W$ boson pair in $pp$ collisions at $\sqrt{s} = 8$ TeV with the ATLAS detector,''
  JHEP {\bf 1601}, 032 (2016)
  doi:10.1007/JHEP01(2016)032
  [arXiv:1509.00389 [hep-ex]].
  %%CITATION = doi:10.1007/JHEP01(2016)032;%%
  
\bibitem{Aad:2015kna} 
  G.~Aad {\it et al.} [ATLAS Collaboration],
  %``Search for an additional, heavy Higgs boson in the $H\rightarrow ZZ$ decay channel at $\sqrt{s} = 8\;\text{ TeV }$ in $pp$ collision data with the ATLAS detector,''
  Eur.\ Phys.\ J.\ C {\bf 76}, no. 1, 45 (2016)
  doi:10.1140/epjc/s10052-015-3820-z
  [arXiv:1507.05930 [hep-ex]].
  %%CITATION = doi:10.1140/epjc/s10052-015-3820-z;%%
  
\bibitem{Aad:2014fha} 
  G.~Aad {\it et al.} [ATLAS Collaboration],
  %``Search for new resonances in $W\gamma$ and $Z\gamma$ final states in $pp$ collisions at $\sqrt s=8$ TeV with the ATLAS detector,''
  Phys.\ Lett.\ B {\bf 738}, 428 (2014)
  doi:10.1016/j.physletb.2014.10.002
  [arXiv:1407.8150 [hep-ex]].
  %%CITATION = doi:10.1016/j.physletb.2014.10.002;%%
  
\bibitem{Aad:2015fna} 
  G.~Aad {\it et al.} [ATLAS Collaboration],
  %``A search for $ t\overline{t} $ resonances using lepton-plus-jets events in proton-proton collisions at $ \sqrt{s}=8 $ TeV with the ATLAS detector,''
  JHEP {\bf 1508}, 148 (2015)
  doi:10.1007/JHEP08(2015)148
  [arXiv:1505.07018 [hep-ex]].
  %%CITATION = doi:10.1007/JHEP08(2015)148;%%

\bibitem{CMS:2014eda} 
  CMS Collaboration [CMS Collaboration],
  %``Search for di-Higgs resonances decaying to 4 bottom quarks,''
  CMS-PAS-HIG-14-013.
  %%CITATION = CMS-PAS-HIG-14-013;%%

\bibitem{Cheung:2015cug} 
  K.~Cheung, P.~Ko, J.~S.~Lee, J.~Park and P.~Y.~Tseng,
  %``A Higgcision study on the 750 GeV Di-photon Resonance and 125 GeV SM Higgs boson with the Higgs-Singlet Mixing,''
  arXiv:1512.07853 [hep-ph].
  %%CITATION = ARXIV:1512.07853;%%
  
\bibitem{D'Eramo:2016mgv} 
  F.~D'Eramo, J.~de Vries and P.~Panci,
  %``A 750 GeV Portal: LHC Phenomenology and Dark Matter Candidates,''
  JHEP {\bf 1605}, 089 (2016)
  doi:10.1007/JHEP05(2016)089
  [arXiv:1601.01571 [hep-ph]].
  %%CITATION = doi:10.1007/JHEP05(2016)089;%%
  
\bibitem{Carmi:2012in} 
  D.~Carmi, A.~Falkowski, E.~Kuflik, T.~Volansky and J.~Zupan,
  %``Higgs After the Discovery: A Status Report,''
  JHEP {\bf 1210}, 196 (2012)
  doi:10.1007/JHEP10(2012)196
  [arXiv:1207.1718 [hep-ph]].
  %%CITATION = doi:10.1007/JHEP10(2012)196;%%

\bibitem{Cheung:2015dta} 
  K.~Cheung, P.~Ko, J.~S.~Lee and P.~Y.~Tseng,
  %``Bounds on Higgs-Portal models from the LHC Higgs data,''
  JHEP {\bf 1510}, 057 (2015)
  doi:10.1007/JHEP10(2015)057
  [arXiv:1507.06158 [hep-ph]].
  %%CITATION = doi:10.1007/JHEP10(2015)057;%%

\bibitem{CMS:2015kwa} 
  The ATLAS and CMS Collaborations,
  %``Measurements of the Higgs boson production and decay rates and constraints on its couplings from a combined ATLAS and CMS analysis of the LHC pp collision data at $\sqrt{s}$ = 7 and 8 TeV,''
  ATLAS-CONF-2015-044, 
  %%CITATION = ATLAS-CONF-2015-044;%%
  %``Measurements of the Higgs boson production and decay rates and constraints on its couplings from a combined ATLAS and CMS analysis of the LHC pp collision data at sqrt s = 7 and 8 TeV,''
  CMS-PAS-HIG-15-002.
  %%CITATION = CMS-PAS-HIG-15-002;%%

\end{thebibliography}
\end{document}